# Emerging Non-Volatile Opto-electronic Resistive-Memories for Next-Generation Photonic Integrated Circuits

**Santosh Kumar[1, *], Mukesh Kumar[2], Eunso Shin[1], Bassem Tossoun[3], Stanley Cheung[1, #]**

[1] *Department of Electrical and Computer Engineering, North Carolina State University, Raleigh, NC. 27606, USA*
[2] *Department of Electrical Engineering and Centre of Advanced Electronics, Indian Institute of Technology Indore, Indore, 453552, India*
[3] *Hewlett Packard Labs, Large-Scale Integrated Photonics Laboratory, 820 N. McCarthy Blvd., Milpitas, CA. 95035, USA*
* *skumar59@ncsu.edu*
[#] *scheung3@ncsu.edu*

***ABSTRACT: -*** Photonic integrated circuits have emerged as a powerful platform for high-speed communication, sensing, and information processing due to their large bandwidth, low latency, and inherent parallelism. However, the absence of efficient, scalable, and non-volatile memory elements remains a fundamental limitation for realizing fully programmable and adaptive photonic systems. Conventional electronic memories introduce significant energy overhead, latency, and architectural inefficiencies due to repeated optical–electrical conversions. Non-volatile opto-electronic resistive memories (OERMs) have recently emerged as a promising solution to address these challenges by integrating memory functionality directly within the photonic domain. These devices combine resistive switching mechanisms with optical readout, enabling persistent state retention, multilevel programmability, and energy-efficient operation. In this review, we provide a comprehensive overview of OERMs, spanning from fundamental physical mechanisms to system-level applications. We first discuss the underlying resistive switching phenomena, including filamentary conduction, interface-type switching, phase-change transitions, and ionic migration, with particular emphasis on their interaction with confined optical modes. We then examine key material platforms such as metal oxides, transparent conducting oxides, phase-change materials, and emerging two-dimensional systems, highlighting their performance trade-offs. Furthermore, we analyse device architectures and benchmark their performance in terms of switching energy, speed, endurance, and optical modulation efficiency. The integration of OERMs into programmable photonic circuits, neuromorphic systems, and in-memory optical computing architectures is critically discussed. Finally, we outline the major challenges and future research directions toward scalable, reliable, and intelligent photonic systems. OERMs represent a transformative technology that bridges the gap between memory and photonics, paving the way for next-generation reconfigurable and energy-efficient optical computing platforms.

**TABLE OF CONTENTS**

## 1. Introduction

Photonic integrated circuits (PICs) have been demonstrated as a versatile platform for high-speed communications, sophisticated sensing, and information processing due to their high bandwidth, ultralow latency, and inherent parallelism through wavelength multipl[1–3]. Nevertheless, most photonic devices remain passive or allow only volatile tuning, inherently constraining the post-fabrication programmability and adaptability of PICs.

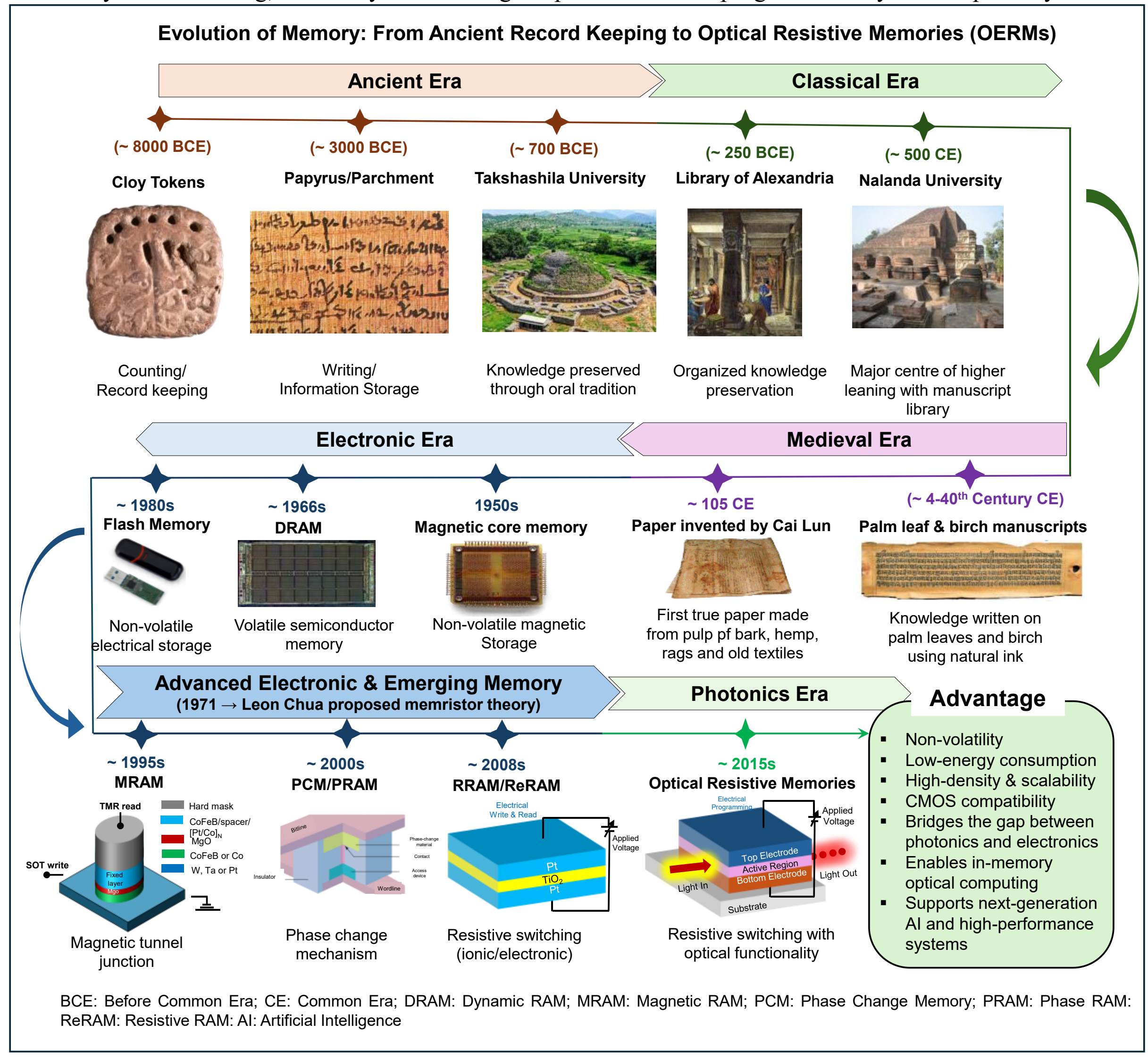

**Fig. 1.** Evolution of memory technologies from early human knowledge preservation systems to modern and emerging non-volatile electronic and photonic memories. The timeline highlights the transition from ancient

storage media, including papyrus and manuscript-based systems, to electronic memory technologies such as MRAM, PCM, and RRAM, culminating in the recent emergence of OERMs. This progression illustrates the continuous advancement toward higher storage density, energy efficiency, and functional integration, ultimately enabling programmable and intelligent photonic systems.

To realize the full potential of future systems-spanning neuromorphic computing and optical interconnects through to quantum information processing-there is an immediate need for embedded, reconfigurable, and non-volatile memory devices that are inherently photonic platform compatible[3,4].

Conventional electronic volatile memories like static random-access memory (SRAM), and dynamic random-access memory (DRAM) need constant power to hold their states and are thus unsuitable for ultralow-power or energy-efficient photonic systems[4]. Although advanced non-volatile memories like flash or resistive random-access memory (ReRAM) encounter significant challenges when combined with photonic circuits, including losses in optical-electrical conversions, speed mismatches, bandwidth bottlenecks, and parasitic wiring overheads[5,6]. These problems limit the achievement of fully scalable, programmable, and energy-efficient photonic architectures. The continuous evolution of memory technologies, from early human knowledge preservation systems to modern electronic and emerging photonic memories, reflects the increasing demand for higher storage density, faster operation, improved energy efficiency, and enhanced functional integration. As illustrated in Fig. 1, memory technologies have progressed from conventional electronic memories toward advanced non-volatile platforms including MRAM, PCM, RRAM, and more recently opto-electronic resistive memories (OERMs), highlighting the growing convergence between memory functionality and integrated photonics[6–9]. Non-volatile opto-electronic resistive memories (OERMs) represent a revolutionary class of devices that integrate resistive switching and non-volatile state retention directly into the photonic domain[10–13]. By enabling electrical or optical programming of states, along with persistent optical readout, OERMs provide several key capabilities:

- Reconfigurability: - allowing post-fabrication tuning of filters, resonators, and phase shifters for adaptive and programmable circuits.
- Scalability: - reducing electrical interconnect requirements while supporting dense device integration on photonic platforms.
- Energy efficiency: - offering zero static-power retention and ultra-low write energies suitable for sustainable large-scale systems.
- High speed: - achieving nanosecond-scale switching, compatible with high-bandwidth communication and computing.
- Multilevel storage: - OERMs can support multiple stable resistance/optical states within a single device, enabling higher information density and weight encoding for neuromorphic and in-memory computing.
- Hybrid electrical–optical control: - The ability to program electrically and read optically (or vice versa) makes OERMs versatile for co-integration with both CMOS electronics and photonics.
- Non-volatility and retention stability: - Retained states can persist for months to years without refresh, crucial for long-term programmable circuits and memory.
- Low footprint and CMOS compatibility: - OERMs can be fabricated using scalable nanofabrication processes and are compatible with silicon photonics platforms.
- Endurance and reliability: - Emerging prototypes demonstrate endurance over millions of cycles, which is essential for practical deployment in reconfigurable photonic systems.
- Broad application versatility: - Beyond memory, OERMs can act as tunable phase shifters, optical switches, synaptic weights in neuromorphic networks, or security primitives for encryption.

Recent progress in non-volatile OERMs has demonstrated their capability to go far beyond simple binary switching[10–14]. These devices can support multilevel storage, where multiple stable resistance or optical states are encoded within a single element, significantly increasing the information density. Their low programming energies often in the picojoule to femtojoule range make them highly attractive for energy-constrained systems[10,11]. In addition, nanosecond-scale switching speeds have been reported, ensuring compatibility with the high-bandwidth demands of modern optical communication and processing. Equally important, OERMs exhibit long retention times and excellent non-volatility, allowing optical states to remain fixed for extended periods without continuous power supply. As conceptually illustrated in Fig. 2, OERMs combine electrically controlled resistive switching with optical readout and can be seamlessly integrated within photonic integrated circuits to enable a broad range of functionalities including non-volatile optical memory, computing, communication systems, sensing, photodetection, and spectroscopy. Their ability to directly modulate and retain optical states within integrated nanophotonic platforms makes OERMs highly promising for future programmable and intelligent photonic systems.

Beyond these device-level merits, OERMs open new opportunities at the system level. Their ability to retain and modulate optical states directly within a PIC enables reconfigurable photonic circuits, where filters, resonators, and phase shifters can be programmed and locked without dynamic power consumption[14–17]. In computing applications, OERMs serve as the foundation for photonic in-memory computing, reducing the need for repeated optical-to-electrical conversions and minimizing data movement bottlenecks. The possibility of encoding weighted connections in multilevel states makes them highly suitable for neuromorphic photonics, where massive parallelism and adaptive learning can be realized at the speed of light. Furthermore, OERMs are being explored for photonic tensor operations, which are critical for machine learning and artificial intelligence workloads[12,15,18]. OERMs are also being explored for adaptive optical communication systems, programmable optical routing networks, optical FPGA architectures, and secure photonic hardware platforms[14,16,17].

This review offers an in-depth overview of the non-volatile opto-electronic resistive memory state-of-the-art. The discussion starts with device physics and material engineering approaches to successful resistive switching in optically addressable materials. It continues to examine scalable integration methods and fabrication routes compatible with high-volume photonic platforms. Finally, the review overview looks at emerging applications and system-level implications, emphasizing how OERMs are leading the way toward reconfigurable, smart, and energy-efficient photonic integrated circuits.

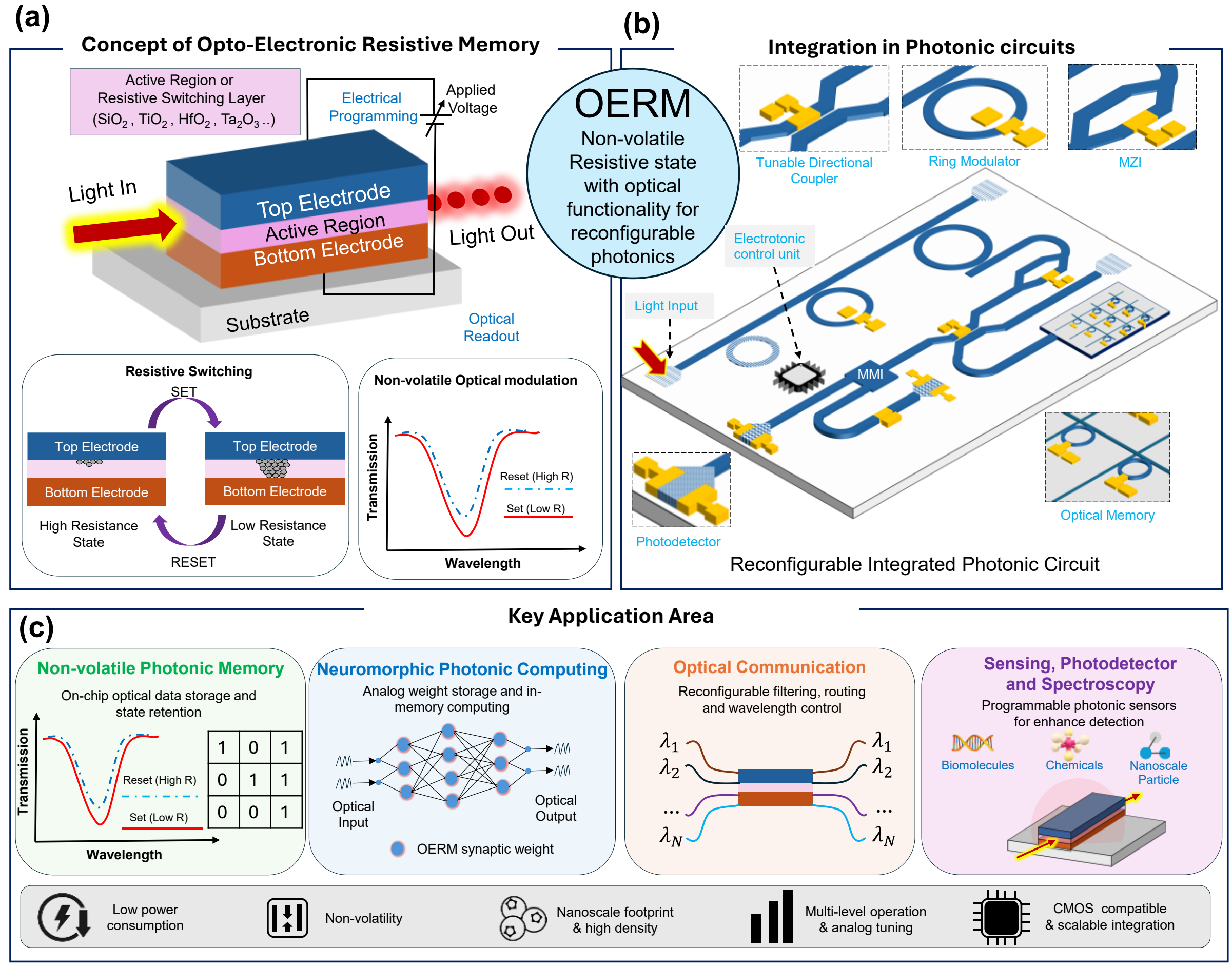

**Fig. 2.** Conceptual illustration of opto-electronic resistive memories (OERMs), showing electrically controlled resistive switching with optical readout, their integration within photonic integrated circuits, and their potential applications in memory, computing, communication systems, sensing, photodetector and spectroscopy.

## 2. Fundamentals, Light-Matter Interaction, and Material Platforms of Opto-Electronic Resistive Memories

The emergence of OERMs stems from the convergence of two historically distinct research directions: resistive switching memories in electronics and photonic integrated circuits for information transmission and processing. Resistive switching was initially developed in the electronic memory community as an attractive alternative to charge-based storage technologies, primarily because of its simple device structure, scalability, low programming voltage, and non-volatility. In parallel, photonic integrated circuits evolved as a powerful platform for high-bandwidth and low-latency signal manipulation, but their practical deployment in large-scale adaptive systems has long been constrained by the absence of compact and efficient non-volatile memory elements. OERMs bridge this gap by integrating programmable resistance states with optical functionality, thereby enabling persistent modulation, memory retention, and reconfigurable photonic operation within a single device platform.

Unlike conventional electronic resistive memories, OERMs must satisfy an additional requirement: the physical process responsible for resistive switching must also induce a sufficiently strong and repeatable change in the optical response of the device. This means that the switching region cannot be treated merely as an electrical element; instead, it must be carefully positioned within or near the optical mode so that any material reconfiguration affects absorption, refractive index, scattering, or phase accumulation. Therefore, the performance of OERMs is not governed solely by electrical switching characteristics such as set/reset voltage or endurance, but also by optical parameters including extinction ratio, insertion loss, modulation depth, spectral response, and compatibility with integrated photonic architectures. As a result, understanding OERMs requires a unified discussion of switching physics, light–matter interaction, and material platforms.

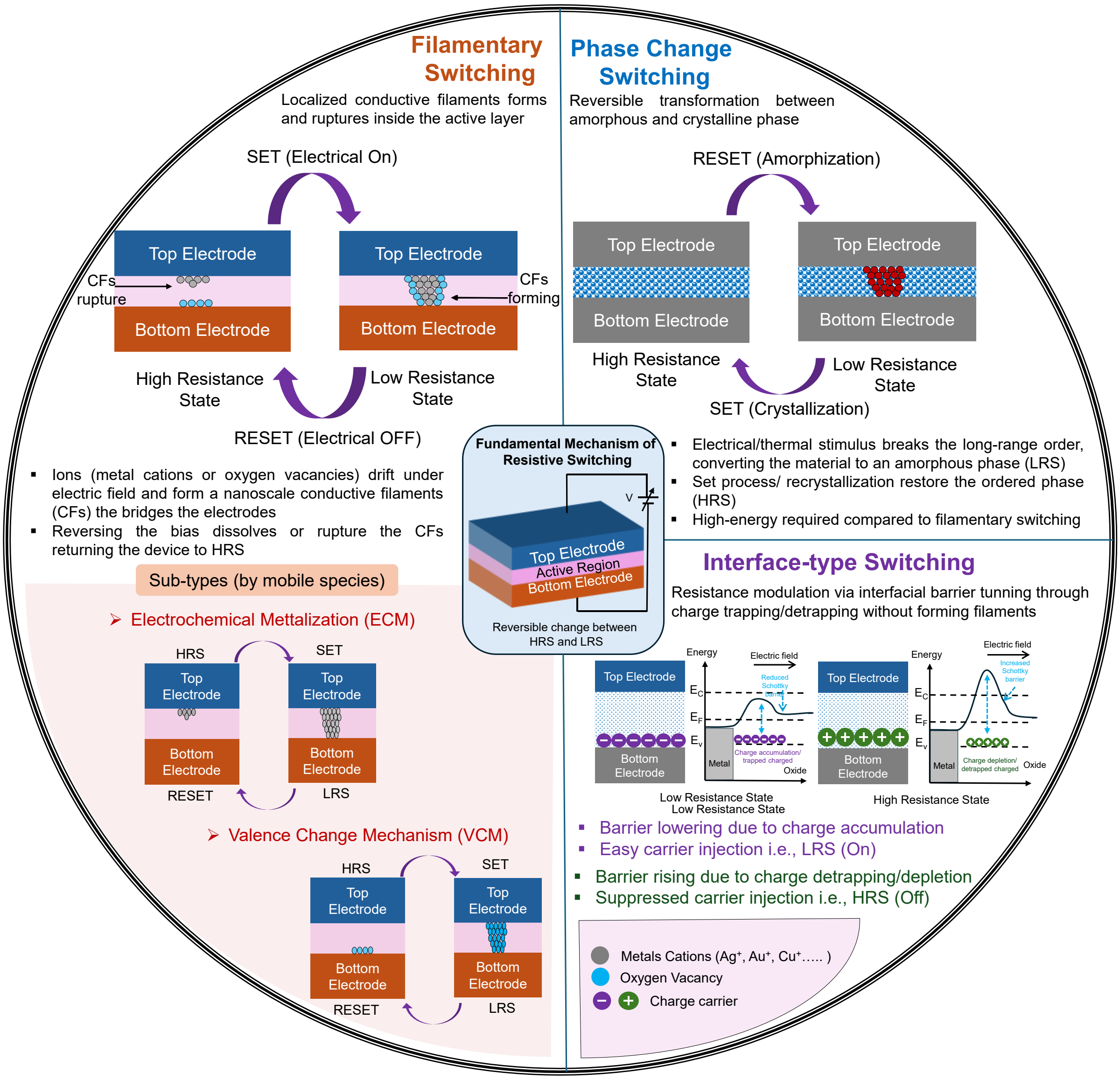

**Fig. 3.** Comprehensive overview of the fundamental mechanisms of resistive switching, illustrating filamentary, phase-change, and interface-type switching within a unified framework. Filamentary switching relies on the formation and rupture of localized conductive filaments driven by ionic migration, while phase-change switching is governed by reversible transitions between amorphous and crystalline states in chalcogenide materials. In contrast, interface-type switching originates from modulation of interfacial energy barriers through charge trapping/detrapping and defect redistribution, without the formation of conductive filaments. The central schematic highlights the common metal–insulator–metal (MIM) device structure and the underlying role of ionic transport in all mechanisms.

## 2.1 Fundamental of Resistive Switching

Resistive switching refers to the reversible modulation of electrical resistance in a material under the action of an external stimulus, most commonly an applied electric field[5,6,9]. In its simplest form, a resistive memory device toggles between a high-resistance state (HRS) and a low-resistance state (LRS), which can be interpreted as logical states for binary storage. In more advanced cases, intermediate states can also be stabilized, allowing analog or multilevel operation suitable for neuromorphic and in-memory computing applications[9,19]. These programmable states remain retained even after the removal of power, making resistive memories particularly attractive for energy-efficient non-volatile operation. At the device level, resistive switching typically takes place in a nanoscale active layer sandwiched between two electrodes, in a in a metal–insulator–metal (MIM) configuration, as schematically illustrated in Fig. 3.[6]. The microscopic origin of switching can involve ionic migration, conductive filament formation, oxygen vacancy redistribution, interface redox reactions, charge trapping/detrapping, or structural phase transformations[5,9,19,20]. Although these mechanisms were initially investigated for purely electronic non-volatile memories, they have become increasingly important in photonics because the same nanoscale reconfiguration processes can simultaneously alter optical properties such as absorption, refractive index, scattering, and phase accumulation when integrated with optical waveguides, resonators, or plasmonic structures[10–12].

One of the most extensively studied switching mechanisms is filamentary switching, where an externally applied electric field drives the formation of a conductive filament through an otherwise insulating medium[5,19,21,22]. As shown in Fig. 3, the filament may consist of active metal ions such as Ag or Cu in electrochemical metallization (ECM) memories, or oxygen vacancy chains in valence-change memories (VCMs)[19,20,23–25]. During the SET operation, the conductive filament progressively bridges the electrodes, resulting in a substantial reduction in electrical resistance. Conversely, during the RESET process, the filament is partially or completely ruptured, restoring the high-resistance state. Filamentary switching is particularly attractive because it enables low-voltage operation, fast switching dynamics, nanoscale scalability, and high integration density. However, the stochastic nature of filament formation and rupture can introduce device-to-device variability, cycle-to-cycle fluctuations, and localized Joule heating, all of which may affect the reproducibility and long-term stability of optical modulation in OERMs[20].

A second major category is interface-type switching, where resistance modulation originates from gradual electronic and ionic modifications at the interface between adjacent materials rather than from the formation of a localized conductive filament[26,27]. These resistance variations may arise from charge trapping/detrapping, defect redistribution, modulation of Schottky barriers, or interfacial redox reactions Resistive switching in transition metal oxides[26]. Compared with filamentary switching, interface-type mechanisms generally exhibit more spatially distributed switching behavior, smoother analog tunability, and improved switching uniformity. Such characteristics are highly desirable for photonic synapses, multilevel optical memory, and analog photonic computing applications, where stable intermediate optical states and continuous weight tuning are required. Nevertheless, interface-type switching may exhibit lower resistance contrast and slower switching speed compared with highly localized filamentary mechanisms [26–28].

Another important class of resistive switching relies on reversible structural transformations between distinct material phases, commonly referred to as *phase-change switching* [8,29]. In these devices, the active material reversibly transitions between amorphous and crystalline states, each possessing significantly different electrical conductivity and optical constants. Phase-change materials (PCMs), particularly chalcogenide compounds such as Ge2Sb2Te5 (GST), have therefore become highly important for non-volatile photonics because they simultaneously provide strong electrical contrast and large refractive index modulation[8,30–33]. As illustrated in Fig. 3, phase-change switching can induce substantial modification of optical transmission, reflection, and phase response, enabling deterministic and broadband optical programmability. Compare with filamentary and interface-type switching, phase-change operation often offers improved multilevel programmability. However, it typically requires higher programming energy due to thermally induced crystallization and amorphization processes and may introduce thermal crosstalk in densely integrated photonic systems[8,29,34,35].

In many resistive memory platforms, ionic migration and electrochemical effects fundamentally govern the switching behavior, either directly or indirectly[5,36,37]. Under an applied bias, mobile ionic species such as metal cations or oxygen vacancies redistribute within the active region, thereby modifying local stoichiometry, conductivity, and defect concentration. In electrochemical metallization cells, active metal ions dissolve from the electrode, migrate through the dielectric, and reduce to form conductive metallic filaments[19]. In valence-change memories, oxygen vacancy migration alters the local electronic structure and conductivity of the oxide layer[20]. These ionic processes become particularly significant in nanoscale photonic devices because strong optical confinement, localized electric fields, and Joule heating can further accelerate and spatially localize ionic transport. Consequently, ionic dynamics strongly influence not only the electrical switching characteristics but also the magnitude, repeatability, and long-term stability of the optical modulation response.

A critical distinction between purely electronic resistive memories and OERMs lies in the requirement for optical accessibility of the switched region[10–12]. In conventional electronic memories, a conductive filament formed anywhere within the active layer is sufficient if it modifies electrical conductivity. In OERMs, however, the switched region must spatially overlap with the optical mode profile to produce meaningful optical modulation. If the conductive filament, interfacial modification, or phase-transformed region remains outside the optical field distribution, the electrical resistance may change without inducing a substantial optical response. Therefore, the performance of OERMs must be evaluated through the combined perspective of electrical switching dynamics and light–matter interaction. This dual requirement makes the integration of resistive switching mechanisms with waveguide, resonator, slot-waveguide, and plasmonic photonic structures particularly important for achieving efficient and scalable non-volatile photonic functionality.

## 2.2 Light-Matter Interaction in OERMs

The defining feature of OERMs is the direct coupling between a non-volatile resistive switching process and a guided optical mode within an integrated photonic environment[10,11]. This coupling transforms a conventional resistive memory element into an active photonic component capable of storing and modulating optical signals. Consequently, the efficiency of OERMs is governed not only by electrical switching characteristics but also by the strength and nature of light-matter interaction. As illustrated in Fig.4. (a), the propagating optical mode interacts directly with the electrically programmable nanoscale switching region, enabling modulation of the transmitted optical signal through electrically induced material reconfiguration. Optical modulation in OERMs originates from switching-induced changes in local material properties, including refractive index, absorption coefficient, free-carrier concentration, and scattering behavior[13,14,38]. The resulting optical response is determined by three key factors: (i) the intrinsic optical contrast of the material, (ii) the spatial overlap between the optical mode and the switching region, and (iii) the degree of optical field confinement[11,14,20]. Among these, mode overlap and confinement play a particularly critical role, as even large material changes may yield weak modulation if they occur outside the optical field, whereas small perturbations can produce strong effects when located within high-field regions.

In conventional dielectric waveguides, the optical mode is relatively delocalized, which limits interaction with nanoscale switching regions[39–41]. To overcome this limitation, device architectures such as slot waveguides, plasmonic waveguides, and hybrid plasmonic structures have been developed. These configurations enable strong field localization within nanoscale regions, significantly enhancing light- matter interaction[42–46]. These architectures spatially compress the optical field into deeply subwavelength regions, thereby increasing the interaction between the propagating optical mode and the active switching material. As schematically illustrated in Fig. 4(b), strong optical mode confinement within the switching region enables substantial modulation of optical transmission between the HRS and LRS. In particular, slot and hybrid plasmonic geometries allow the switching medium to be positioned within regions of maximum electric field intensity, thereby maximizing modulation efficiency within a compact footprint[43,47–50].

Transparent conducting oxides, especially indium tin oxide, are particularly well-suited for such architectures due to their tunable carrier concentration and operation near the epsilon-near-zero regime[51–57]. In this regime, small variations in carrier density can induce large changes in optical permittivity, enabling efficient modulation. When combined with strong optical confinement, TCO-based OERMs can achieve high modulation depth with reduced device dimensions and low switching energy[13,51]. However, enhanced confinement introduces an inherent trade-off between modulation efficiency and optical loss. Although plasmonic and hybrid structures significantly enhance light-matter interaction, they often suffer from increased propagation loss due to metal absorption and scattering. Therefore, device design requires careful optimization to balance extinction ratio, insertion loss, and energy efficiency. Additionally, the stochastic nature of resistive switching, particularly in filamentary systems, introduces variability in the spatial location of the active region [9,20]. Since optical modes are spatially non-uniform, this variability can lead to fluctuations in optical response, even when electrical characteristics appear consistent. Electrically induced material reconfiguration can modulate optical phase, resonance wavelength, coupling strength, polarization response, and spectral characteristics depending on the photonic architecture employed. For example, in resonator-based systems, switching-induced refractive index changes can shift resonance wavelengths and modify quality factors, while in interferometric structures such as Mach-Zehnder interferometers (MZIs), they can induce persistent non-volatile phase shifts. Furthermore, bistable optical transmission and hysteresis behavior, as conceptually illustrated in Fig. 4(c), demonstrate the capability of OERMs to realize non-volatile and reconfigurable photonic operation. These functionalities highlight that OERMs extend far beyond conventional memory applications and can serve as fundamental building blocks for programmable photonic processors, optical neural networks, reconfigurable routing systems, and energy-efficient photonic computing platforms[14,18,31,58,59].

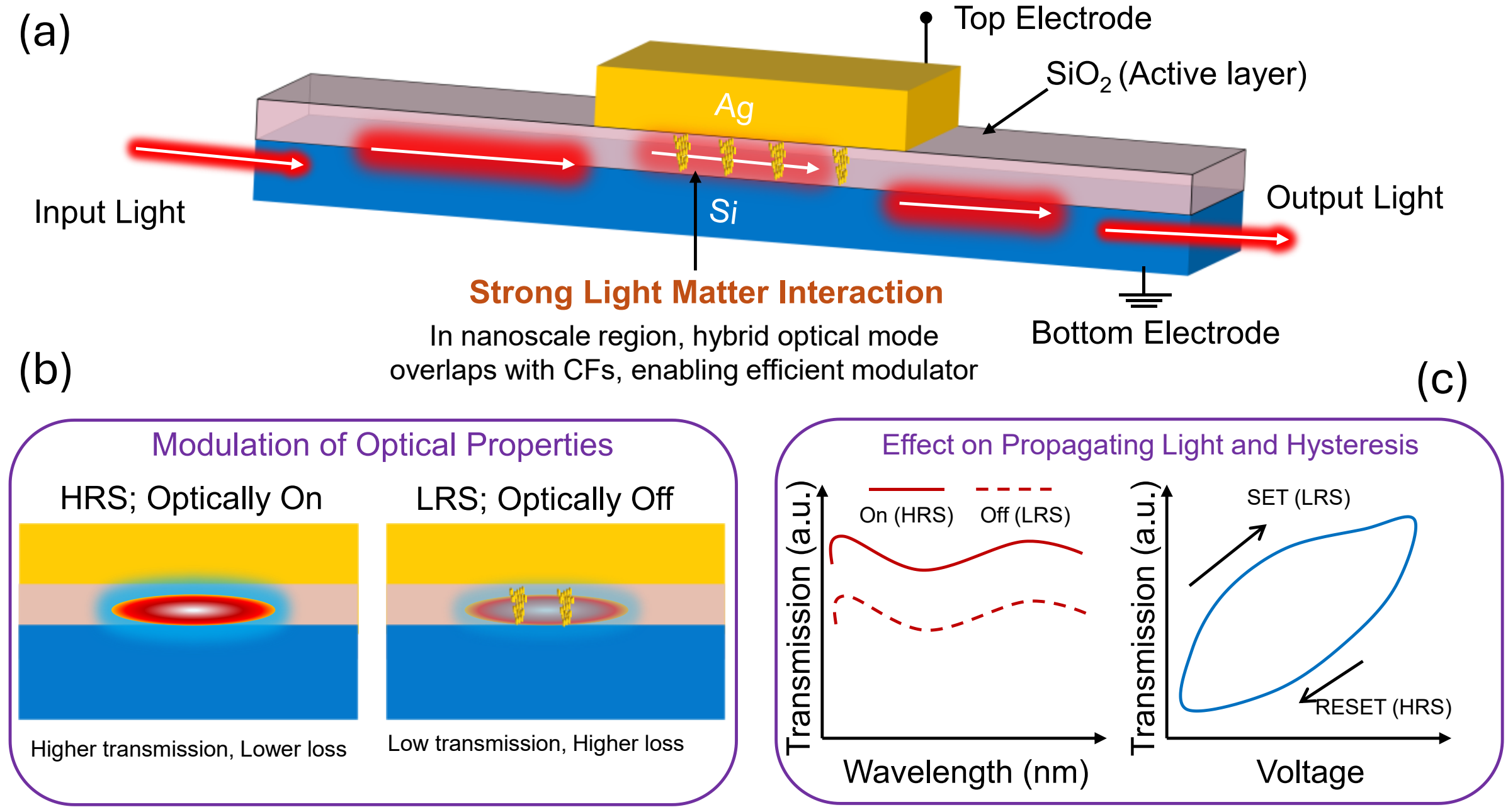

**Fig. 4.** (a) Schematic illustration of light propagation through the device, highlighting its interaction with the nanoscale active switching region under an applied electrical bias. (b) Light–matter interaction in an opto-electronic resistive memory (OERM), showing strong optical mode confinement within the switching region and the resulting modulation of optical properties between the high-resistance state (HRS) and low-resistance state (LRS). (c) Variation of optical transmission as a function of wavelength for ON and OFF states, along with the corresponding optical hysteresis behavior, demonstrating non-volatile and bistable photonic functionality.

## 2.3 Material Platforms for OERMs

The choice of material platform plays a decisive role in determining the electrical, optical, and integration characteristics of OERMs. An ideal material for OERM applications should support reliable non-volatile switching, exhibit appreciable optical contrast between programmed states, operate at low energy, maintain long retention, withstand repeated cycling, and be compatible with established nanofabrication processes. In practice, no single material system satisfies all these requirements simultaneously, and the field is shaped by trade-offs among switching speed, endurance, optical loss, compactness, and manufacturability. Different material platforms including metal oxides, TCOs, (PCMs, two-dimensional materials, and organic/hybrid systems provide distinct advantages depending on the targeted photonic functionality and switching mechanism[6,21,57,60–63].

***Metal-oxide systems: -*** Metal-oxide materials remain among the most extensively studied platforms for resistive switching, owing to their simple device structure, CMOS compatibility, and well-established switching physics. Systems based on $SiO_2$, $HfO_2$, $TiO_2$, $Al_2O_3$, and related oxides have demonstrated filamentary or interface-type switching with good non-volatility and relatively low operating voltages. In OERMs, metal-oxide systems are attractive because they can be integrated into silicon photonic structures and can support large resistance contrast through oxygen vacancy migration or electrochemical metallization. In particular, silver- or copper-based oxide memories are highly relevant because metallic filament formation can produce pronounced local changes in absorption and scattering when placed inside an optical mode. Such devices are promising for compact optical switches, memory units, and non-volatile attenuators. However, metal filament formation may also increase optical loss, especially if the metal resides directly in a high-field optical region. Therefore, the geometry must be carefully engineered to maximize switching-induced modulation without incurring excessive baseline absorption. Transition-metal oxides such as $TiO_2$ and $HfO_2$ offer advantages in terms of process maturity and robust switching, but their direct optical functionality may be weaker unless paired with optical confinement strategies that amplify interaction with the modified region. Their role in OERMs is therefore often strongest when integrated into slot waveguides, resonators, or hybrid structures where even localized defect-driven changes become optically significant. Recent demonstrations of light-assisted electro-metallization memories, waveguide-integrated ReRAMs, and filament-controlled photonic switching further establish oxide-based systems as one of the foundational platforms for OERMs [6,9,21,64].

***Transparent conducting oxides: -*** Transparent conducting oxides, including indium tin oxide and aluminum-doped zinc oxide, have emerged as particularly promising materials for OERMs because they sit at the intersection of electronics and photonics. Electrically, they can support carrier modulation and resistive-state evolution. Optically, they offer tunable refractive index and absorption, especially near the epsilon-near-zero regime. This dual functionality makes TCOs highly suitable for compact photonic memory and modulation devices. Among the TCO family, ITO is especially important due to its compatibility with integrated photonics and its strong optical tunability under carrier redistribution. When incorporated into hybrid plasmonic or slot waveguide configurations, ITO can produce substantial optical response over extremely small footprints. This is one of the main reasons why ITO-based OERMs are considered strong candidates for dense programmable photonic circuits. They can, in principle, combine low switching energy, non-volatility, strong mode overlap, and subwavelength confinement in a single platform. At the same time, TCO-based devices present fabrication and stability challenges. Their optical constants are highly sensitive to deposition conditions, stoichiometry, annealing, and interface quality. Small process variations can significantly shift the operating point, especially in ENZ-based designs. In addition, the trade-off between optical confinement and loss remains significant. Thus, although TCOs offer exceptional promise for ultracompact OERMs, their practical deployment requires tight process control and careful co-design of material properties with optical geometry. The emergence of ENZ-assisted optical modulation, hybrid plasmonic waveguides, and vertically coupled ITO resonator structures has significantly accelerated research toward low-energy and ultracompact non-volatile photonic devices[13,51–53,55,65–67].

***Phase-change materials: -*** Phase-change materials such as $Ge_2Sb_2Te_5$ and related chalcogenide compounds are perhaps the most established non-volatile optical switching materials in integrated photonics. Their appeal lies in the large contrast in refractive index and absorption between amorphous and crystalline phases, which can be exploited for strong non-volatile optical modulation. Phase-change materials have been widely investigated in waveguide-integrated memory cells, resonators, and reconfigurable photonic processors. In the broader OERM landscape, phase-change materials represent a valuable benchmark and, in some contexts, a competing non-volatile technology. They excel in achieving deterministic optical contrast and broadband operation. However, their switching is often thermally driven, which can require relatively high programming energy and raise concerns about thermal crosstalk, heater integration, and long-term material fatigue. For large-scale programmable photonic circuits, these issues may limit the density or energy efficiency of phase-change approaches. Compared with filamentary oxide memories or TCO-based electro-optic devices, phase-change materials often offer stronger intrinsic optical contrast but may be less attractive when ultralow-power operation, nanoscale filament control, or direct resistive-state tunability is desired. Therefore, in a review of OERMs, they should be positioned both as a relevant material platform and as a reference point against which other non-volatile photonic memory technologies can be compared. Recent progress in multilevel photonic memories, optical tensor cores, photonic neural networks, and electro-optic memristive switching demonstrates the continuing importance of PCM-assisted photonic systems for programmable and neuromorphic photonics[11,12,18,31,34,68–71].

***Two-dimensional materials and perovskites: -*** Emerging material platforms such as two-dimensional materials and halide perovskites are attracting interest because they introduce new possibilities for ultrathin active regions, strong surface sensitivity, multifunctionality, and heterogeneous integration. Two-dimensional materials can exhibit electrically tunable optical properties, defect-mediated switching, and strong interaction with confined optical modes due to their atomic-scale thickness. Perovskites, on the other hand, offer unusual ionic dynamics and strong optoelectronic response, which may be leveraged for non-volatile optical functionality. These materials are scientifically exciting because they may enable OERMs with extremely small active volumes and novel physical mechanisms beyond those accessible in conventional oxides or chalcogenides. However, they currently face substantial challenges in long-term stability, reproducibility, environmental robustness, and scalable fabrication. For this reason, they are best viewed as emerging rather than mature OERM platforms. Their near-term value lies in expanding the design space and uncovering new switching optical coupling mechanisms, even if large-scale deployment remains a longer-term goal. Graphene, $MoS_2$, $WS_2$, black phosphorus, and van der Waals heterostructures have shown considerable promise for ultracompact electro-optic modulation and nonlinear photonic interaction, motivating increasing interest in low-dimensional OERMs[72–76]. Table 1 summarizes the major material platforms used in OERMs and compares their switching mechanisms, optical modulation capability, energy efficiency, speed, and integration compatibility. Among these, silicon-compatible plasmonic and TCO-assisted structures generally offer the strongest optical confinement and lowest footprint, whereas phase-change materials provide superior retention and deterministic optical contrast. In contrast, two-dimensional and organic/hybrid materials provide unique opportunities for ultrathin and flexible photonic systems but currently face challenges in reliability and large-scale manufacturability

**Table 1:** Comparison of material platforms for OERMs, highlighting differences in switching mechanism, optical modulation, speed, energy efficiency, and integration compatibility across metal-oxide semiconductor, PCM, plasmonic/ITO, 2D, and organic materials[11,12,66–69,71,73–77,14,78–83,16,18,25,34,51,52,65].

| Material Platform | Silicon Hybrid Plasmonic | Phase Change Materials | 2D Materials | Organic/Hybrid Materials |
|---|---|---|---|---|

| Typical Examples | (Ag/Au/Cu) and ($SiO_2$/$TiO_2$/$HfO_2$/ZnO/ITO/$Ta_2O_5$) on Si/$SiN_X$ | GST, $VO_2$, $Ge_2Sb_2TE_2$ | $MoS_2$, $WS_2$, h-BN, Graphene, BP | Polymer, Perovskites, Hybrid oxide |
|---|---|---|---|---|
| **Switching Mechanism** | Filaments | Amorphous ↔ Crystalline | Charge trapping/Interface modulation | Ion migration in organic matrix |
| **Optical Effect** | Strong light-matter interaction | Large Δn, Δk | Strong Change in absorption/Index | Moderate |
| **ER Ratio or On-Off Ratio (in dB)** | 30-50 | 30-40 | 20-40 | 15-30 |
| **Switching Speed** | ps-ns | 100 ns-µs | ~ 1 Gbps | <1 ns |
| **Energy Efficiency** | Very Low | Moderate | Low | Moderate |
| **Integration Compatibility** | High (CMOS compatible) | Moderate | Moderate | Moderate |
| **Key Advantage** | Ultra-compact, scalable, robust, strong confinement, low power | Long retention time | Ultra-thin, high-modulation, low-power | Low-temperature processing, flexible |

**Comparative Perspective on Mechanisms and Materials: -** A critical perspective is essential when evaluating OERMs because no single switching mechanism or material platform is universally optimal. The most suitable choice depends strongly on the target application, whether that application prioritizes ultracompact memory cells, low-loss programmable phase control, analog weight storage, broadband switching, or high-endurance operation. Filamentary oxide-based devices are compelling for compact non-volatile optical memories because they can achieve large electrical contrast and potentially strong optical modulation in highly confined geometries. Their main limitations are variability and possible excess optical loss caused by metallic or vacancy-rich conductive paths. Interface-type devices are more attractive for analog and multilevel applications because they generally offer smoother state evolution, though they may sacrifice switching contrast or speed. Phase-change materials provide large and deterministic optical contrast, making them highly valuable for reconfigurable photonics, but their energy cost and thermal management challenges can be significant. TCO-based platforms, especially those using ITO, occupy a particularly attractive middle ground for advanced integrated photonics because they combine optical tunability, compactness, and compatibility with strong field confinement, although fabrication sensitivity remains a major concern. Moreover, recent demonstrations of programmable photonic processors, photonic tensor accelerators, optical memristors, and integrated photonic neural networks indicate that hybrid material integration will likely dominate future large-scale OERM architectures[13,31,34,68,69,84].

From a broader viewpoint, OERMs should be understood as a co-design problem involving switching physics, material selection, and optical architecture. A material with excellent resistive memory properties may perform poorly in photonics if its optical loss is too high or if it cannot be positioned in the region of strongest optical field. Similarly, a material with large optical tunability may not be suitable for memory if its state retention or endurance is insufficient. The future of OERMs therefore depends not on isolated optimization of switching materials alone, but on integrated engineering of material platforms, nanoscale device geometry, optical mode distribution, and circuit-level requirements. For high-density programmable photonic integrated circuits, the most promising directions are likely to be those that offer a balanced compromise among low switching energy, strong non-volatile optical contrast, low insertion loss, multilevel programmability, and fabrication compatibility. In this regard, hybrid plasmonic TCO-based devices, oxide filament-based slot structures, and carefully engineered waveguide-integrated non-volatile memories stand out as especially important directions for continued research. In parallel, emerging approaches based on optical FPGA architectures, photonic tensor cores, and in-memory optical computing further emphasize the growing importance of multifunctional and scalable OERM material platforms for next-generation intelligent photonic systems[14,16,34,68].

## 3. Device Architectures and Performance Metrics

The realization of OERMs in PICs critically depends on device architecture, which governs the strength of light–matter interaction, switching efficiency, optical loss, and scalability. While the underlying resistive switching mechanisms define the electrical behaviour, the optical performance is primarily dictated by how effectively the switching region interacts with the guided optical mode. Consequently, device architecture plays a central role in translating nanoscale material changes into measurable and controllable optical responses[9,20,85,86]. Over the past decade, multiple device configurations have been explored to integrate resistive switching functionality into photonic platforms. These architectures differ in how the optical mode overlaps with the active switching region and how efficiently electrical programming can be translated into optical modulation. In general, OERM devices can be classified into waveguide-integrated structures, resonator-based devices, interferometric configurations, plasmonic/hybrid plasmonic platforms, and emerging laser-integrated and memristive photonic systems. Each approach offers distinct advantages in terms of footprint, modulation efficiency, and functionality, and also introducing specific trade-offs must be carefully balanced, making them suitable for different application domains ranging from optical memory to neuromorphic computing[87–91].

***Waveguide-Integrated OERMs: -*** Waveguide-integrated OERMs represent the most straightforward approach for embedding memory functionality into photonic circuits. In these devices, the resistive switching material is incorporated within or adjacent to an optical waveguide, enabling direct modulation of the propagating optical signal[92,93]. Early demonstrations employed metal-oxide systems integrated on silicon waveguides, where resistive switching induces changes in absorption or refractive index, leading to modulation of optical transmission[25]. However, in conventional dielectric waveguides, the optical field is relatively delocalized, resulting in weak interaction with nanoscale switching regions. Consequently, large material changes or long interaction lengths are often required to achieve significant modulation. To address this limitation, slot waveguide architectures have been widely adopted. In such configurations, the optical field is strongly confined within a nanoscale low-index region, where the switching material is placed. This enhances mode overlap and enables efficient modulation with reduced footprint [11,63]. For instance, Ag/$SiO_2$ and $HfO_2$-based resistive switching layers integrated in slot waveguides have demonstrated non-volatile optical modulation with improved extinction ratios and reduced device length[12,18,81,94]. Representative waveguide-integrated OERM architectures are illustrated in Fig. 5 (a–d), highlighting the diversity of material platforms and optical confinement strategies employed to realize non-volatile photonic functionality. Figure 5(a) demonstrates a plasmonic electro-optical synapse based on an Au–$HfO_2$–Ti/Au slot waveguide architecture, where strong gap-plasmon confinement enhances localized heating and oxygen-vacancy-mediated filament formation for synaptic and memory operation[23]. Figure 5(b) presents a hybrid $VO_2$–Si optical memory integrated on a rib waveguide platform, in which the electrically induced insulator-to-metal transition of $VO_2$ modulates optical transmission and enables non-volatile photonic memory functionality[95]. Figure 5(c) illustrates a wavelength-scale hybrid Si–$VO_2$ electro-intensity at subwavelength scales. *Adapted from [65].* (d) Optical memristor based on thin-film PZT integrated in a ridge waveguide. Cross-ponding mode profile distribution for TE mode. SEM image of the fabricated memristor. Representation of before and after poling process of PZT unit cell. These architectures highlight the role of material integration, optical confinement, and resistive or phase-change mechanisms in enabling compact, non-volatile photonic memory and modulation functionalities. Waveguide based OERMs. *Adapted from [96], licensed under CC BY 4.0* absorption waveguide device exhibiting distinct ON/OFF optical propagation characteristics arising from the phase-dependent optical constants of $VO_2$ [65]. More recently, ferroelectric waveguide-integrated optical memristors based on PZT thin films, as shown in Fig. 5(d), demonstrated programmable non-volatile optical modulation through electrically controlled ferroelectric domain switching and electro-optic tuning within integrated ridge waveguides[96]. These studies collectively demonstrate how waveguide geometry, optical confinement, and material engineering directly influence the efficiency, compactness, and programmability of OERMs. More recently, integration with transparent conducting oxides such as ITO has further enhanced performance. ITO-based waveguide OERMs exploit carrier modulation and ENZ effects to achieve strong optical response with low energy consumption. Foundational studies by Alam et al. and Krasnok et al. showed that ENZ materials enable strong optical modulation with minimal material variation, making them highly attractive for compact OERMs[49,52,67,97,98]. In addition, hybrid plasmonic and slot-assisted ENZ architectures provide enhanced overlap between the optical mode and active switching region, thereby enabling ultracompact non-volatile photonic devices with improved modulation efficiency and reduced switching energy[99,100]. These devices are particularly attractive for compact, energy-efficient photonic memory elements. Despite these advantages, waveguide-integrated OERMs must carefully balance modulation efficiency and propagation loss, especially when lossy materials or metallic filaments are introduced into the optical path. Therefore, optimization of waveguide geometry, active-material placement, and optical mode engineering remains critical for achieving high extinction ratio, low insertion loss, multilevel programmability, and scalable photonic integration in future OERM platforms[51,101,102].

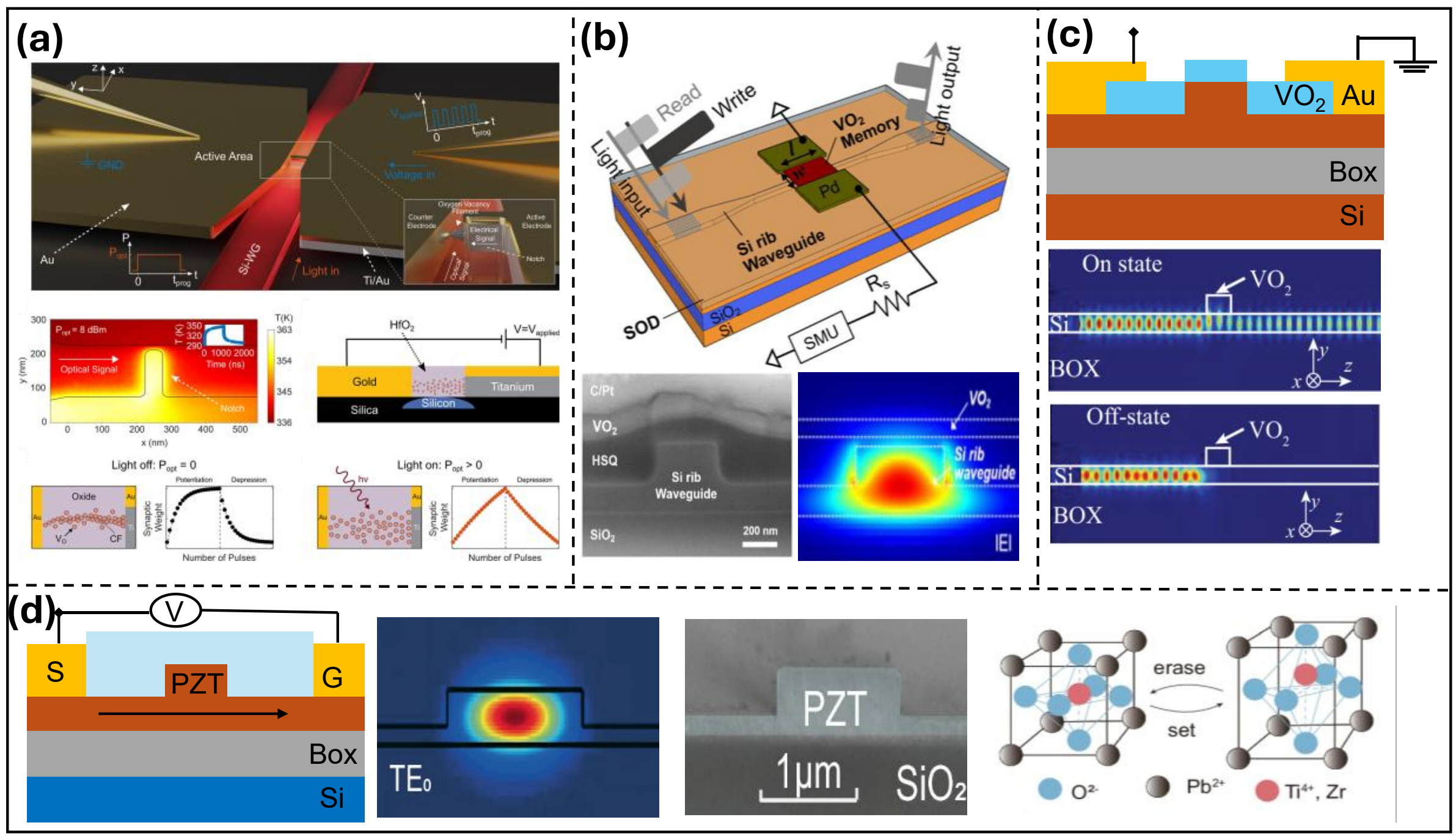

**Fig. 5. Waveguide-based OERMs illustrating diverse device architectures and underlying switching mechanisms. (a)** Nanoscale electro-optical synapse based on a plasmonic Au-$HfO_2$-Ti/Au slot waveguide with a nanoscale notch, where incident optical power is converted into a highly confined gap plasmon, inducing localized heating. Resistive switching occurs through the formation and rupture of oxygen-vacancy-based conductive filaments within the $HfO_2$ layer. The notch region enhances optical field confinement and thermal localization, enabling efficient electro-optical modulation and synaptic weight tuning. *Adapted with permission[23] from American Chemical Society.* **(b)** Waveguide-integrated $VO_2$-Si optical memory device, where the phase transition of $VO_2$ modulates the optical mode propagating in the silicon waveguide. The structure enables non-volatile optical memory through electrically driven changes in optical absorption and refractive index, with strong optical mode overlap in the active region. *Adapted from [95], licensed under CC BY 4.0* **(c)** Hybrid Si-$VO_2$ waveguide electro-absorption device demonstrating state-dependent optical transmission. The electric field distribution and propagation characteristics differ significantly between ON and OFF states, resulting in modulation of optical

***Resonator-Based OERMs: -*** Resonator-based OERMs, including microring and microdisk resonators, leverage optical resonance to enhance sensitivity to material changes. In these devices, even a small variation in refractive index or absorption induced by resistive switching can result in a significant shift in resonance wavelength or modulation depth[57,80]. By integrating the switching material at high-field regions of the resonator, strong non-volatile tuning can be achieved with minimal energy consumption. Phase-change materials have played a pivotal role in the domain. In a landmark study, Zhang et. al. demonstrated non-volatile using GST integrated microring resonator[103], while extended the concept towards the neuromorphic photonic system capable of parallel information processing[52,104,105]. More recently, electrically programmable resonator-based OERMs employing memristive switching mechanisms have emerged as promising alternatives to thermally driven phase-change systems. As shown in Fig. 6(a), heterogeneously integrated III–V/Si photonic memresonators combine optical resonance with embedded oxide-based memristive switching layers, where different resistive states directly modulate the resonance condition and transmission spectra, thereby enabling compact non-volatile photonic memory with low static power consumption[80]. Similarly, racetrack resonators integrated with embedded ReRAM switching layers, shown in Fig. 6(b), exploit strong evanescent-field interaction between the guided optical mode and the active switching region to achieve programmable high- and low-transmission states suitable for non-volatile optical memory operation[57]. Beyond individual resonators, optical memristor arrays based on memresonator architectures, illustrated in Fig. 6(c), highlight the potential of extending conventional electrical memristor crossbar concepts toward wavelength-parallel photonic computing and optical matrix-vector multiplication systems[106]. These works established resonator-based architectures as a powerful platform for non-volatile photonics. More recently, resistive switching mechanisms have been integrated into resonator configurations, offering electrically programmable alternatives to thermally driven phase-change systems. Hybrid plasmonic ring resonators and Si-ITO-based devices have demonstrated enhanced tuning efficiency and strong light–matter interaction[15,107]. An emerging concept is the development of mem-resonator architectures, where the resistive state directly controls the resonance condition. These devices combine non-volatility with resonant enhancement, enabling compact and energy-efficient photonic memory elements[69]. However, they are inherently wavelength-selective and sensitive to fabrication variations. Furthermore, switching-induced variability can lead to resonance instability, which is particularly critical for dense photonic circuits.

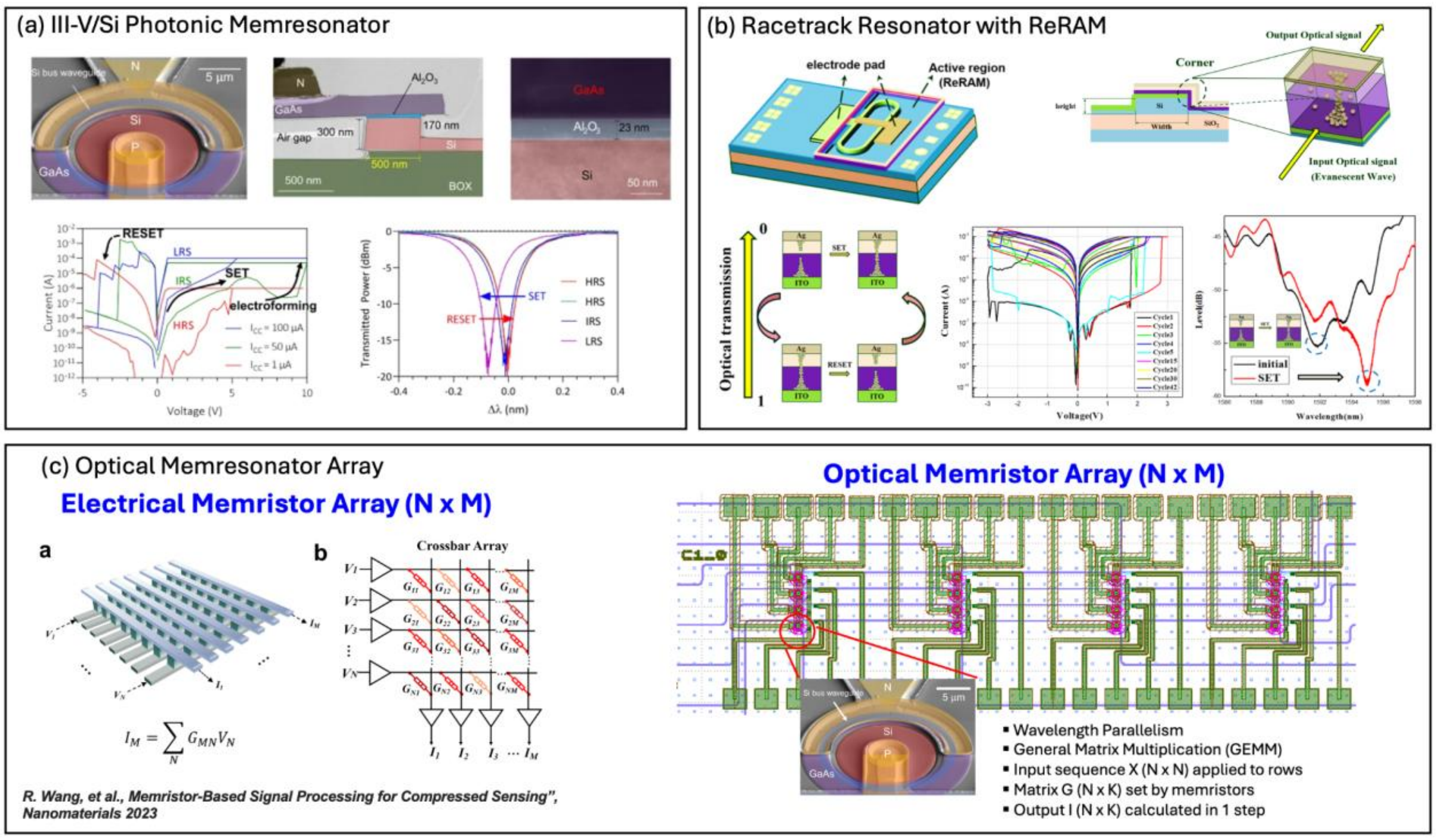

**Fig.6. Resonator based OERMs. (a)** Memristive III–V-on-silicon microring resonator (memresonator) showing device structure, SEM/TEM cross-sections, and memristive switching behavior. Different resistance states, controlled by current compliance, modulate the optical resonance, resulting in distinct transmission spectra and enabling non-volatile photonic memory. *Adapted from [80], licensed under CC BY 4.0* **(b)** MMI-integrated racetrack resonator with embedded ReRAM (Ag/$Al_2O_3$/BFO/ITO) demonstrating non-volatile optical memory. Conductive filament formation (SET) and rupture (RESET) modulate optical absorption and refractive index, leading to low and high transmission states, respectively, with strong evanescent field interaction in the active region. *Adapted from [57].* (c) Left: Electrical memristor crossbar array (M × N) enabling matrix-vector multiplication (MVM), where input voltages are mapped through the conductance matrix to generate output currents based on Ohm's and Kirchhoff's laws. Right: Optical memristor array based on III-V silicon memresonator illustrating the extension of such architectures into photonic systems, enabling parallel optical signal processing and integration of memory and computation Left image:- *Adapted form [106] licensed under CC BY 4.0.*

***Interferometric OERMs (Mach–Zehnder Architectures): -*** Mach-Zehnder interferometers (MZIs) provide a robust platform for implementing OERMs with phase-sensitive operation. In this configuration, the switching material is embedded in one arm of the interferometer, and resistive switching induces a change in optical phase or amplitude, resulting in modulation at the output[78]. MZI-based OERMs offer several advantages, including broadband operation, reduced wavelength sensitivity, relatively low insertion loss, and linear response characteristics. These features make them particularly suitable for large-scale photonic circuits which are essential for analog photonic computing and neuromorphic applications, where continuous tuning and multi-level states are required[34]. Recent studies have demonstrated non-volatile phase modulation using resistive switching materials integrated into silicon MZIs, achieving stable multi-level optical states and low-energy operation[18]. Notably, Cheung et al. demonstrated memristor-integrated Mach-Zehnder interferometers on heterogeneous III–V/Si platforms, enabling electrically programmable and non-volatile phase control[34,80]. Figure 7(a) shows a III–V/Si memristive MZI architecture exhibiting electrically programmable optical switching and filtering functionalities through embedded memristive elements. Recently, ferroelectric photonic memristors based on lead zirconate titanate (PZT) have also emerged as promising interferometric OERMs[78]. Figure 7(b) illustrates a PZT-based 2 × 2 optical MZI memristor capable of non-volatile phase modulation through electrically controlled ferroelectric domain poling[96]. Such devices demonstrate stable wavelength tuning, programmable optical states, and reconfigurable transmission characteristics suitable for photonic memory and switching applications. MZI-based memristive photonics has further enabled large-scale programmable photonic computing architectures. Figure 7(c) presents a heterogeneous III–V/Si optical neural network (ONN) platform employing MZI mesh networks integrated with photonic memristive phase shifters and nonlinear activation units for energy-efficient matrix-vector multiplication and AI/ML acceleration[108]. In addition, scalable reconfigurable optical routing has been demonstrated using PZT memristive MZI switch matrices. Figure 7(d) shows a programmable 4 × 4 optical switching architecture employing cascaded PZT optical memristors for dynamically configurable signal routing and non-volatile photonic interconnects[96]. Compared with resonator-based devices, MZIs are less sensitive to wavelength variations, but they typically require a larger footprint due to longer interaction lengths. These devices represent an important step toward scalable programmable photonic processors.

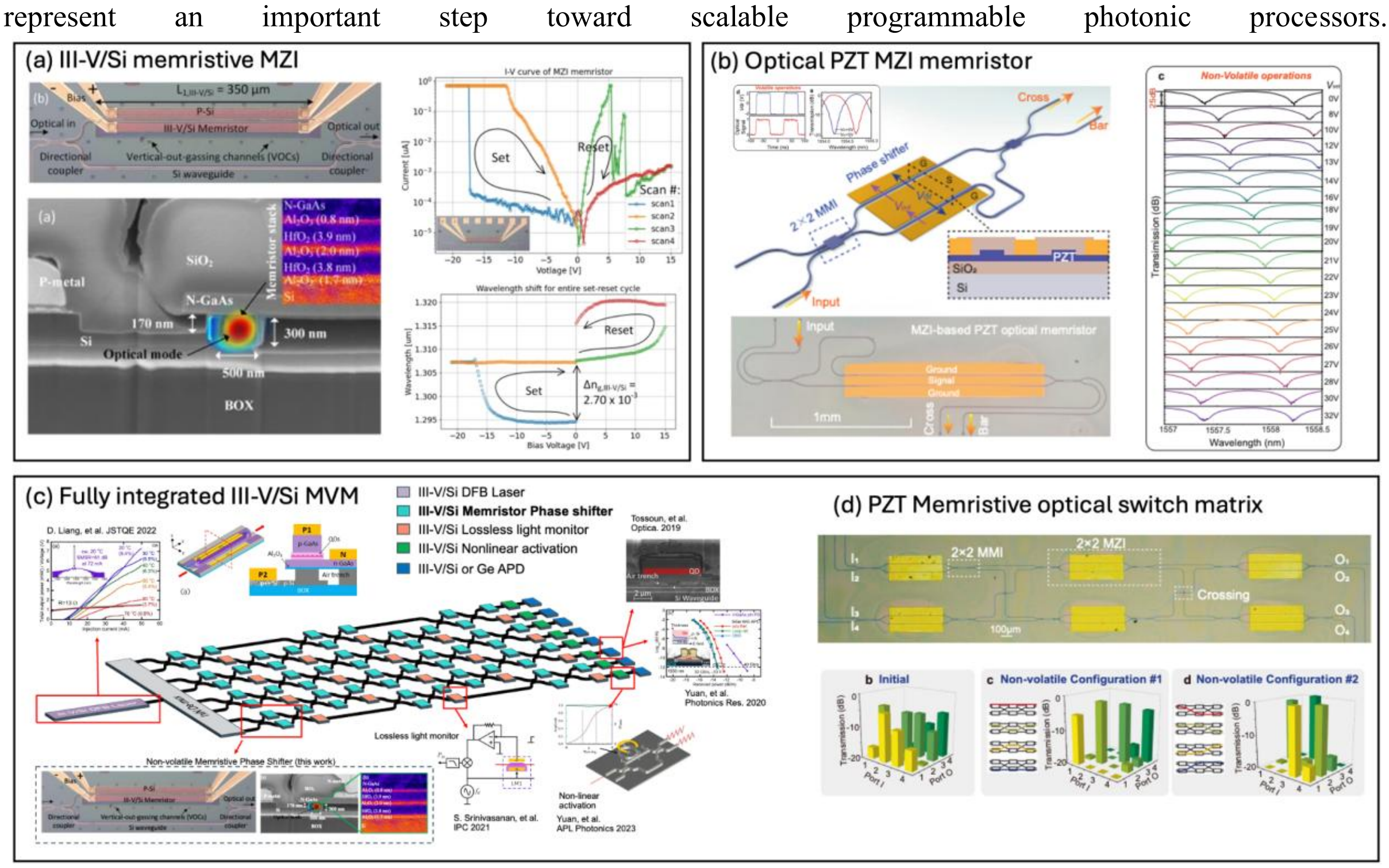

**Fig. 7. MZI-Based OERMs: (a)** Electrically programmable III–V/Si memristor-integrated MZI demonstrating non-volatile optical switching and filtering. SEM cross-sections, optical mode profiles, and electro-optical measurements reveal hysteretic memristive behavior and tunable resonance control. *Adapted from [78], licensed under CC BY 4.0* (**b)** Ferroelectric PZT-based 2 × 2 MZI optical memristor enabling non-volatile phase modulation through electrically controlled domain poling. Optical transmission measurements demonstrate stable wavelength tuning and high-speed optical modulation. *Adapted from [96], licensed under CC BY 4.0* **(c)** Large-scale optical neural network (ONN) architecture based on MZI mesh networks on a heterogeneous III–V/Si platform, enabling programmable matrix–vector multiplication and energy-efficient photonic AI/ML processing. *Adapted from [108], licensed under CC BY 4.0* **(d)** Reconfigurable MZI photonic circuits enabled by scalable PZT optical memristors, demonstrating programmable 4 × 4 optical switching and dynamically configurable routing between multiple input and output ports. *Adapted from [96], licensed under CC BY 4.0*

***Plasmonic and Hybrid Plasmonic OERMs: -*** Plasmonic and hybrid plasmonic architectures enable strong subwavelength confinement of optical fields, significantly enhancing light–matter interaction in nanoscale switching regions[10,109]. This capability is particularly critical for OERMs, where the active switching region is typically confined to nanometre-scale dimensions. Hybrid plasmonic waveguides, which confine light within nanoscale dielectric gaps between metal and semiconductor layers, provide an effective balance between strong confinement and manageable propagation loss. Foundational works by Oulton et al. established the feasibility of such structures for ultra-compact photonic devices[80,110], while subsequent studies have explored their integration into functional nanophotonic systems. Fig. 8(a) illustrates an ultra-compact silicon nanophotonic modulator based on a plasmonic MOS configuration, where strong optical confinement within the ITO layer enhances electro-absorption modulation efficiency[51]. In the context of optically accessible resistive switching, plasmonic confinement has been extensively exploited to enhance switching efficiency. Fig. 8(b) presents a nanoscale plasmonic memristor with optical readout functionality, demonstrating voltage-dependent optical transmission changes associated with resistive switching states[10]. Recent nanophotonic implementations based on engineered metal–insulator–semiconductor structures have demonstrated strong field localization within the switching region, enabling high extinction ratio and low-voltage operation. Figure 8(c) shows a plasmonic memristor operating as a latching optical switch, where conductive filament formation within the metal–insulator–metal stack perturbs the hybrid plasmonic mode and enables non-volatile optical modulation[110]. Hybrid plasmonic slot-based configurations have further enabled optically accessible resistive switching with enhanced endurance and compact footprint.

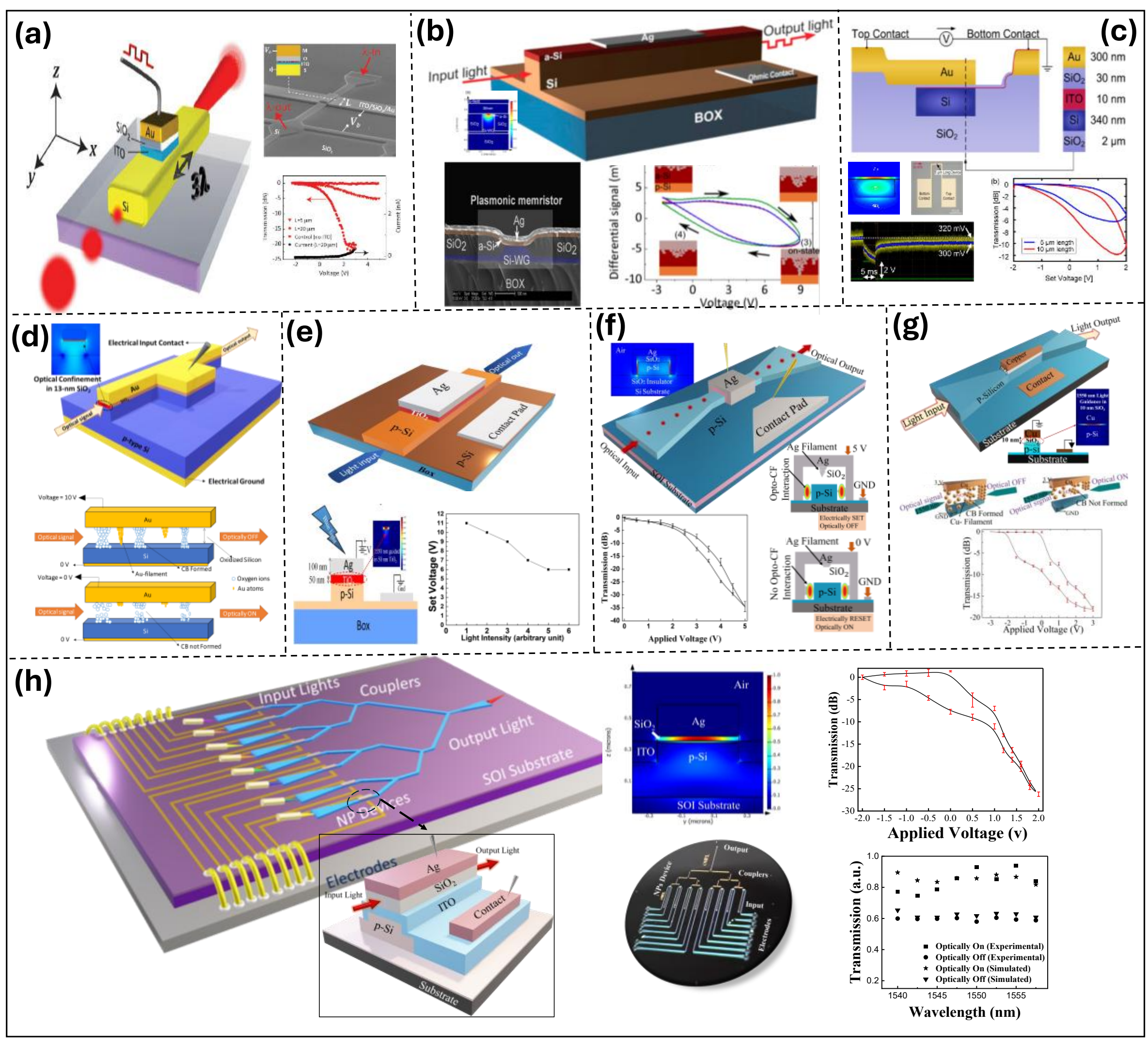

**Fig. 8. Plasmonic and Hybrid Plasmonic (a)** Ultra-compact silicon nanophotonic modulator based on a plasmonic MOS configuration, illustrating strong field confinement in the ITO layer that enhances light–matter interaction and enables high-efficiency electro absorption-based modulation *Adapted from [51].* **(b)** Optically readable plasmonic RRAM showing device architecture and cross-sectional structure, with voltage-dependent optical response demonstrating resistive switching with optical readout. *Adapted from [10], with permission from American Chemical Society*. **(c)** Plasmonic memristor operating as a latching optical switch, where conductive filament formation in the MIM stack perturbs the hybrid plasmonic mode, enabling non-volatile optical modulation *Adapted from [110].* **(d)** Electrically writable silicon nanophotonic resistive memory with optical readout, highlighting hybrid plasmonic confinement and modulation induced by metal filament formation within the dielectric layer. *Adapted from [25].* **(e)** Light-assisted electro-metallization resistive switch, demonstrating optical control of switching dynamics, where increased optical pump intensity reduces the SET voltage. *Adapted from [63].* **(f)** Double-slot hybrid plasmonic waveguide-based resistive switch exhibiting strong subwavelength mode confinement and reliable optical switching through filament formation and rupture. *Adapted from [81]* **(g)** Tapered copper–silicon nanophotonic resistive switch with enhanced modal confinement, showing filament-driven resistive switching and corresponding voltage-controlled optical transmission. *Adapted from [83]* **(h)** Reconfigurable multiwavelength nanophotonic circuit integrating engineered plasmonic resistive switches, enabling electrically controlled optical routing and multiplexed signal processing across multiple channels. *Adapted from [67]*

Figure 8(d) depicts an electrically writable silicon nanophotonic resistive memory employing hybrid plasmonic confinement for non-volatile optical modulation through filament formation inside the dielectric region[25]. Figure 8(e) demonstrates light-assisted electro-metallization in an optically accessible resistive switch, where increased optical pump intensity reduces the SET voltage and enhances switching dynamics[63]. More recently, advanced plasmonic nanophotonic resistive switches have been developed using engineered slot and tapered architectures. Figure 8(f) shows a double-slot hybrid plasmonic waveguide-based resistive switch exhibiting strong subwavelength optical confinement and reliable optical switching characteristics enabled through conductive filament formation and rupture[81]. Figure 8(g) illustrates a tapered Cu–Si nanophotonic resistive switch with enhanced modal confinement, enabling low-power operation and high extinction ratio through voltage-controlled filamentary switching[83]. Similarly, Cu–ITO plasmonic absorber structures have shown enhanced optical modulation with low-voltage tuning and strong extinction characteristics, demonstrating the synergy between plasmonic confinement and carrier-induced modulation[111]. The integration of transparent conducting oxides such as ITO further enhances device performance through epsilon-near-zero (ENZ) effects. As demonstrated by Alam et al., ENZ materials enable strong optical modulation through small carrier variations, which, when combined with plasmonic confinement, results in highly efficient and compact OERMs[52]. Figure 8(h) presents a reconfigurable multiwavelength nanophotonic circuit integrating engineered plasmonic resistive switches for electrically controlled optical routing and multiplexed signal processing across multiple wavelength channels[67]. Overall, these developments highlight that plasmonic and hybrid plasmonic engineering is a central enabler for high-performance OERMs, directly impacting switching efficiency, extinction ratio, energy efficiency, and device scalability. However, managing optical loss associated with metal absorption remains a key challenge for practical large-scale photonic integration.

***Laser-Integrated OERMs: -*** Laser-integrated OERMs represent a significant advancement toward the realization of fully functional photonic systems with embedded memory and light generation capabilities. Unlike passive architectures, these systems directly integrate resistive switching elements within optical cavities, enabling non-volatile control over lasing characteristics such as emission wavelength, output intensity, and threshold behaviour. In such devices, resistive switching modifies the effective refractive index, gain distribution, or intracavity loss, thereby dynamically tuning the resonance condition of the laser cavity. This enables persistent programmability without continuous electrical bias, which is highly advantageous for low-power photonic systems. Figure 9(a) illustrates a heterogeneously integrated III–V/Si memristive micro-ring laser platform, where a memristive switching layer is embedded within the laser cavity to achieve non-volatile wavelength tuning. The structure combines GaAs-based gain material with a silicon photonic platform and incorporates a memristive $Al_2O_3$ switching region that modulates the optical cavity characteristics through reversible resistance-state transitions. Electrical hysteresis measurements confirm memristive SET/RESET behaviour, while the measured laser spectra demonstrate stable and reversible wavelength tuning associated with different resistance states. Such architectures highlight the possibility of integrating memory, tuning, and light generation functionalities within a single compact photonic device[112]. An emerging extension of this concept is shown in Fig. 9(b), where a quantum-dot III–V/Si distributed feedback (DFB) laser is co-integrated with a memristive switching element. In this architecture, resistive switching enables non-volatile modulation of the effective refractive index within the DFB cavity, resulting in persistent wavelength shifts and programmable laser operation. The corresponding electrical hysteresis and wavelength evolution demonstrate the feasibility of electrically programmable laser memories and adaptive on-chip light sources. Such devices are particularly attractive for reconfigurable optical interconnects, programmable transceivers, wavelength-division multiplexing systems, and photonic computing platforms. Furthermore, emerging studies have explored the integration of resistive switching in distributed feedback (DFB) lasers and microcavity lasers, where local modulation of the refractive index enables wavelength tuning and optical memory functionality. These architectures open new possibilities for self-configurable photonic transmitters, tunable laser arrays, and adaptive optical communication systems. Despite their promising potential, laser-integrated OERMs remain in an early stage of development. Key challenges include thermal stability, integration complexity, and maintaining stable lasing operation under repeated switching cycles. Addressing these issues will be critical for their practical deployment. An emerging and relatively unexplored direction involves integrating OERMs directly into laser cavities, such as DFB lasers or hybrid III–V/Si platforms. In these devices, resistive switching can modulate the gain, loss, or refractive index within the cavity, enabling non-volatile control of lasing properties. Such integration opens new possibilities for programmable light sources, including wavelength tuning, intensity control, and on-chip optical memory within active devices. Although still in early stages, this approach is particularly relevant for advanced photonic systems where tight integration of sources and memory is required [113].

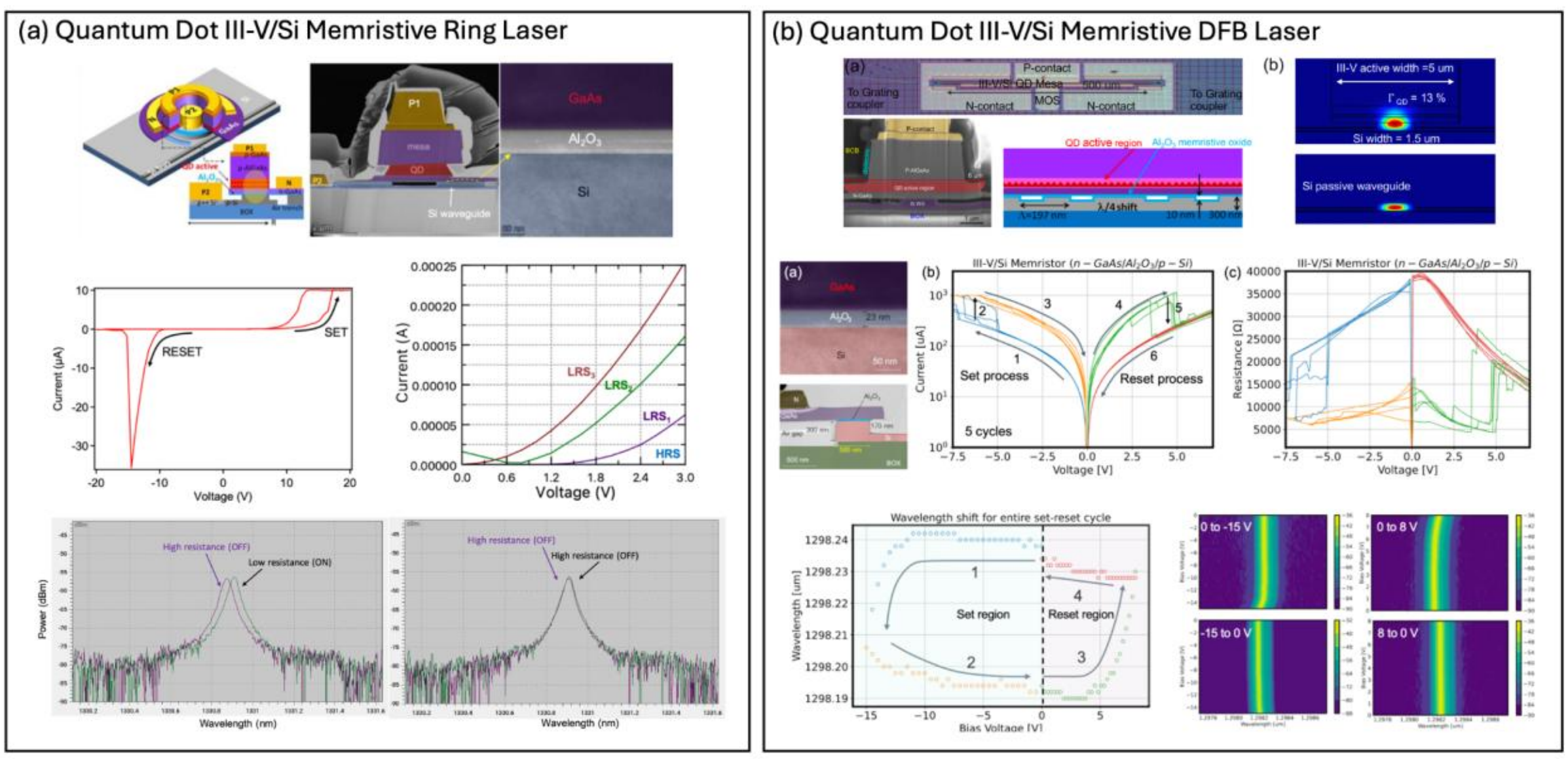

**Fig. 9. Laser Integrated OERMs (a).** Heterogeneously integrated memristive micro-ring laser on silicon with non-volatile wavelength tuning. The figure shows schematic three-dimensional and cross-sectional views of the III-V/Si memristive laser with simulated TE optical mode and active switching region, along with TEM images of the fabricated bonded GaAs-on-Si structure. Electrical characterization demonstrates memristive I–V hysteresis corresponding to reversible switching between high- and low-resistance states, while dynamic measurements and laser output spectra confirm non-volatile wavelength tuning induced by resistance-state modulation. The device demonstrates the potential of laser-integrated OERMs for programmable and energy-efficient photonic memory applications. *Adapted from [112]* (b) Example of non-volatile quantum dot (QD) III-V/Al2O3/ Si distributed feedback (DFB) laser with co-integrated memristor. Electrical hysteresis is shown along with corresponding non-volatile wavelength shifts. *Adapted from [113]*

***Multilevel and Analog OERMs: -*** A defining advantage of OERMs is their inherent ability to support multilevel and analog switching, where multiple stable optical states can be achieved within a single device. This capability arises from the controlled modulation of conductive filaments, interface states, phase configurations, or carrier concentration within the active material. Such multilevel operation is particularly important for neuromorphic photonics, optical signal processing, and in-memory computing, where continuous or discrete intermediate states are required to represent synaptic weights or analog information[11,90,93]. As illustrated in Fig. 10(a), phase-change-material-based nanophotonic memories employing GST integrated on silicon waveguides enable all-photonic non-volatile multilevel storage through intermediate crystallization states, thereby producing multiple distinguishable optical transmission levels[11]. Electrically programmable multilevel photonic random-access memory (P-RAM), shown in Fig. 10(b), further demonstrated high-resolution multibit operation using GSSe-integrated microheater architectures with improved endurance and low-loss operation[34]. Recent nanophotonic resistive switching platforms based on Ag-ITO-$SiO_2$ structures, illustrated in Fig. 10(c), have demonstrated stable multilevel optical states with enhanced optical storage density and controllable analog transmission characteristics [98]. Such devices exploit voltage-controlled filament dynamics and strong plasmonic light confinement to achieve tunable intermediate optical states with long retention and low operating voltage. In addition, phase-change-material-based optical memristive switches integrated in silicon MMI platforms, shown in Fig. 10(d), enable analog optical modulation through partial crystallization of GST, resulting in multiple programmable optical levels [68]. Beyond optical memory, multilevel OERMs have also demonstrated dynamic light modulation and polarization control functionalities. As depicted in Fig. 10(e), filament-based optical memristor systems can simultaneously support multilevel optical storage, analog optical modulation, and wavelength-dependent optical response through electrically controlled resistive switching processes[22]. These developments indicate that OERMs are not limited to binary memory applications but can function as multifunctional photonic elements capable of simultaneously controlling optical amplitude, phase, wavelength, and polarization. In addition to multilevel functionality, OERMs have also been extended toward multiwavelength and spectrally reconfigurable photonic circuits. Reconfigurable nanophotonic systems have demonstrated wavelength-selective modulation and dynamic spectral control, enabling programmable filtering, routing, and multiplexing operations. However, achieving reliable multilevel operation remains challenging due to stochastic filament formation, cycle-to-cycle variability, and device degradation. Future work must focus on improving switching uniformity and developing adaptive control schemes for stable analog photonic operation.

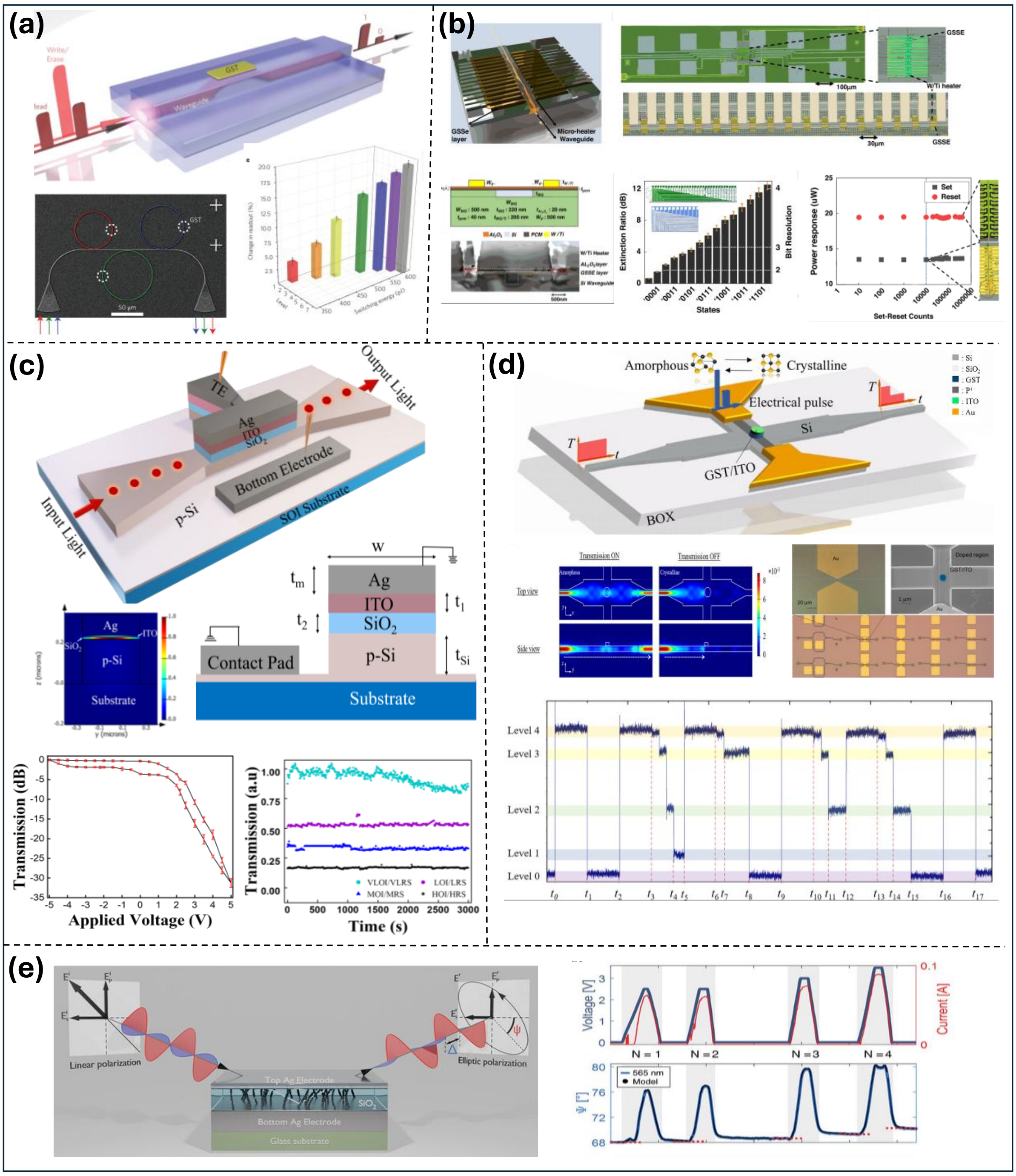

**Fig. 10. Multilevel and Analog OERMs: (a)** All-photonic non-volatile multilevel memory based on phase-change material (GST) integrated on a nanophotonic waveguide, where information is stored in intermediate crystallization states and accessed optically via transmission contrast *Adapted from [11], permission from springer Nature.* **(b)** Electrically programmable multilevel photonic random-access memory (P-RAM) employing GSSe film integrated with microheaters, enabling low-loss, high-resolution multibit operation with improved endurance and cycling stability. *Adapted from [34], licensed under CC BY 4.0* **(c)** Hybrid plasmonic Ag-ITO-$SiO_2$ nanophotonic resistive switch demonstrating multilevel optical states, where voltage-controlled filament dynamics enable stable, tunable transmission levels with long retention. *Adapted from [98].* **(d)** Multilevel optical memristive switch using phase-change material in a silicon MMI platform, where partial crystallization of GST enables analog optical modulation and intermediate switching states. *Adapted from [68], with permission from American Chemical Society*. **(e)** Filament-based optical memristor system enabling multilevel optical storage and dynamic light modulation, where resistive switching governs wavelength-dependent optical response and analog behaviour. *Adapted from [[22]], licensed under CC BY 4.0.*

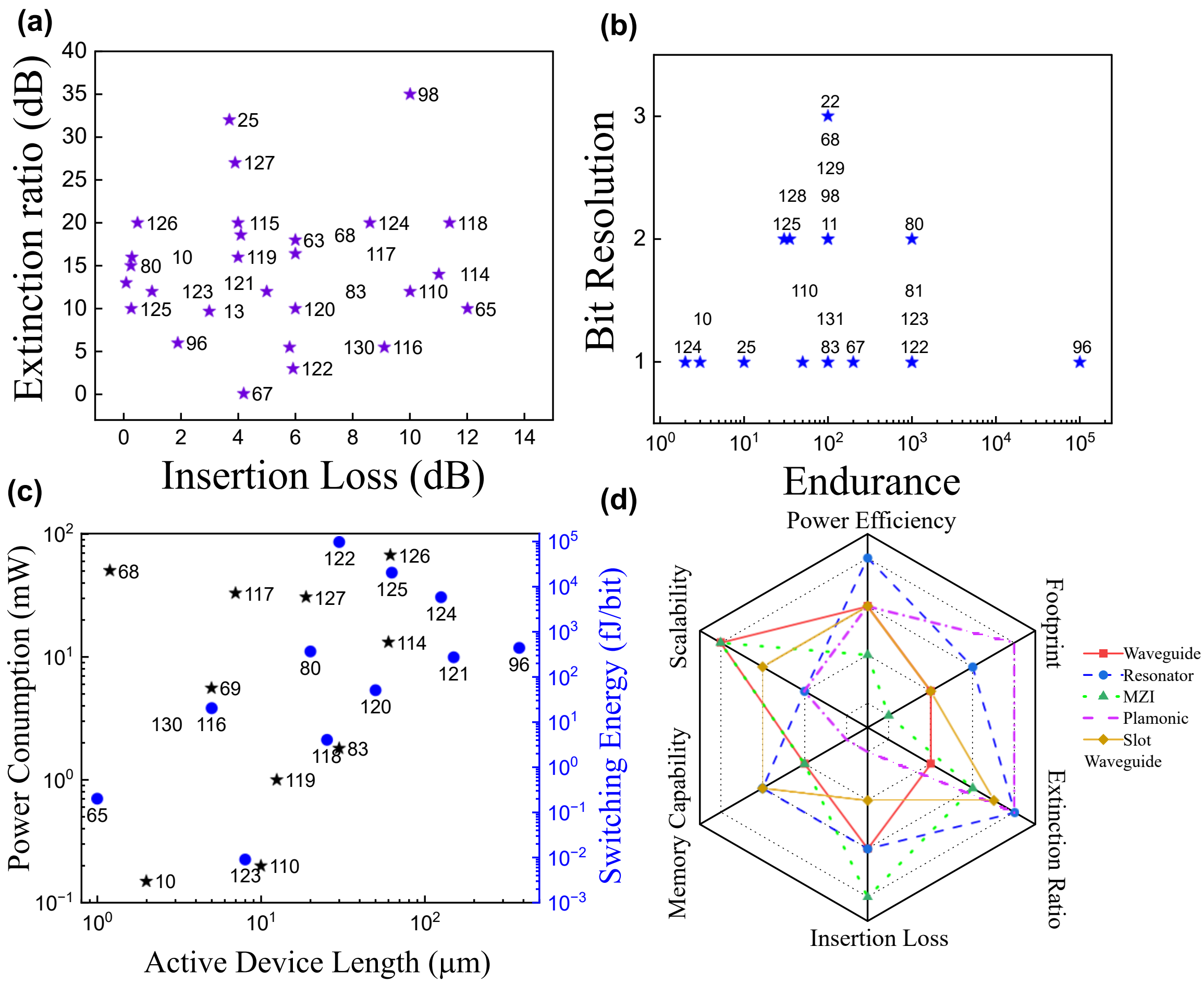

**Fig. 11. Comprehensive benchmarking and comparative analysis of representative OERM platforms and photonic device geometries based on experimentally reported performance** metrics[10,11,81,83,96,98,110,114–118,22,119–128,25,129–131,63,65,67–69,80]**.** (a) Extinction ratio (ER) as a function of insertion loss for various OERM implementations, highlighting the trade-off between modulation depth and optical attenuation across different material and device architectures. (b) Bit resolution versus endurance characteristics, illustrating the relationship between multilevel memory capability and switching reliability for non-volatile photonic memory devices. (c) Power consumption and switching energy as a function of active device length, demonstrating the scaling behavior and energy–footprint trade-offs among diverse OERM configurations. (d) Radar benchmarking of representative OERM photonic architectures, including waveguide-based, resonator-based, MZI based, plasmonic, and slot-waveguide configurations. The radar map compares normalized architecture-level performance trends in terms of power efficiency, footprint, extinction ratio, insertion loss, memory capability, and scalability. Collectively, these comparisons highlight the distinct advantages and limitations of different OERM platforms for integrated photonic memory, programmable photonics, and neuromorphic computing applications.

***Performance Metrics and Benchmarking: -*** The performance of opto-electronic resistive memories (OERMs) is governed by a combination of electrical, optical, memory, and architectural parameters that collectively determine their suitability for integrated and programmable photonic systems. Unlike conventional electronic memories or purely optical modulators, OERMs simultaneously exhibit electrical switching and non-volatile optical modulation, necessitating a multi-dimensional benchmarking framework for comprehensive performance evaluation. Figure 11 summarizes the benchmarking of representative OERM platforms using experimentally reported literature data and normalized architecture-level comparisons.

- *Extinction ratio (ER) and insertion loss:* ER determines the optical modulation depth, while insertion loss reflects signal attenuation introduced by the active device. Benchmarking results reveal a strong trade-off between these parameters, particularly in plasmonic and highly confined structures, where enhanced light–matter interaction enables ER values exceeding 20-30 dB at the expense of increased optical loss, as depicted in Fig. 11(a).

- Bit resolution and endurance: Multilevel optical switching enables multiple stable memory states within a single device, which is highly attractive for high-density optical storage and neuromorphic photonic computing. Reported OERMs demonstrate bit resolutions ranging from binary operation to multilevel states (>2 bits), with endurance spanning from $10^2$ to beyond $10^6$ switching cycles depending on the active material and switching mechanism, as depicted in Fig. 11(b).
- *Power consumption and device footprint:* Device scaling plays a critical role in determining energy efficiency and integration density. Plasmonic and hybrid photonic structures provide ultra-compact footprints due to subwavelength optical confinement, while resonator- and waveguide-based devices offer improved energy efficiency and reduced insertion loss. Benchmarking of active device length versus power consumption highlights the trade-offs between compactness, thermal management, and switching energy, as illustrated in Fig. 11(c).

*Architecture-level benchmarking:* Figure 11(d) shows the radar-map-based comparisons further illustrate the relative strengths and limitations of different OERM photonic geometries, including waveguide-based, resonator-based, MZI-based, plasmonic, and slot-waveguide architectures. Waveguide platforms provide excellent scalability and CMOS compatibility, resonator devices offer high ER and low power operation, MZI architectures enable programmable interference-based functionality with low optical loss, while plasmonic and slot-waveguide devices achieve enhanced optical confinement and compact footprints. These comparisons collectively highlight the diverse design trade-offs involved in optimizing OERMs for programmable photonics, integrated optical memory, and neuromorphic computing applications.

## 4. Pathways to Programmable Photonic Integrated Circuits

The realization of OERMs as practical building blocks for photonic technologies requires their seamless integration into scalable PICs. Beyond individual device demonstrations, this transition demands careful engineering of light-matter interaction, circuit architecture, material platforms, and operational reliability. This section outlines the key pathways enabling OERMs to evolve into programmable and reconfigurable photonic systems.

***Light-Matter Interaction and Mode Engineering: -*** Efficient operation of OERMs relies on strong overlap between the optical mode and the nanoscale switching region. Advanced waveguide geometries, including slot and hybrid plasmonic configurations, enable enhanced field confinement within active regions, thereby improving modulation efficiency and reducing switching energy. Mode engineering techniques such as tapering, adiabatic transitions, and localized field enhancement further optimize coupling into the resistive region. These strategies are critical for achieving high extinction ratio in compact device footprints while maintaining compatibility with integrated photonic platforms.

***Circuit-Level Design and Scalability: -*** Integrating OERMs into PICs requires careful circuit-level design to ensure scalability and performance. Key considerations include device footprint, integration density, and routing of both optical and electrical signals. As device density increases, optical and thermal crosstalk become significant challenges, necessitating optimized layouts and isolation strategies. Additionally, scalable addressing schemes and control architectures are essential for enabling large arrays of programmable elements within reconfigurable photonic circuits.

***Materials and Hybrid Integration Platforms: -*** Material selection plays a pivotal role in determining the performance and scalability of OERM-based PICs. Silicon photonics provides a mature and CMOS-compatible platform, while integration with functional materials such as ITO, metal oxides, and emerging 2D materials enable non-volatile switching and enhanced optical modulation. Heterogeneous integration with III-V materials further extend functionality to active photonic devices, including lasers. Achieving low-loss integration while maintaining fabrication compatibility remains a central challenge in this domain.

***Reliability, Stability, and Endurance: -*** For large-scale deployment, OERMs must exhibit robust performance over repeated switching cycles and extended operation. Key reliability concerns include variability in switching behaviour, thermal effects due to localized heating, and long-term material stability. Device endurance, retention, and optical loss evolution must be carefully optimized through material engineering and structural design. Addressing these challenges is essential for ensuring consistent performance in densely integrated photonic systems.

***Programmability and Reconfigurable Photonic Circuits: -*** The integration of OERMs enables non-volatile programmability in photonic circuits, allowing dynamic control over optical signals without continuous power consumption. Arrays of OERMs can form reconfigurable meshes that support programmable routing, filtering, and phase control. This capability underpins emerging concepts such as optical field-programmable gate arrays (FPGAs) and neuromorphic photonic systems, where OERMs act as tunable weights or memory elements. These developments mark a transition toward adaptive, energy-efficient, and intelligent photonic architectures.

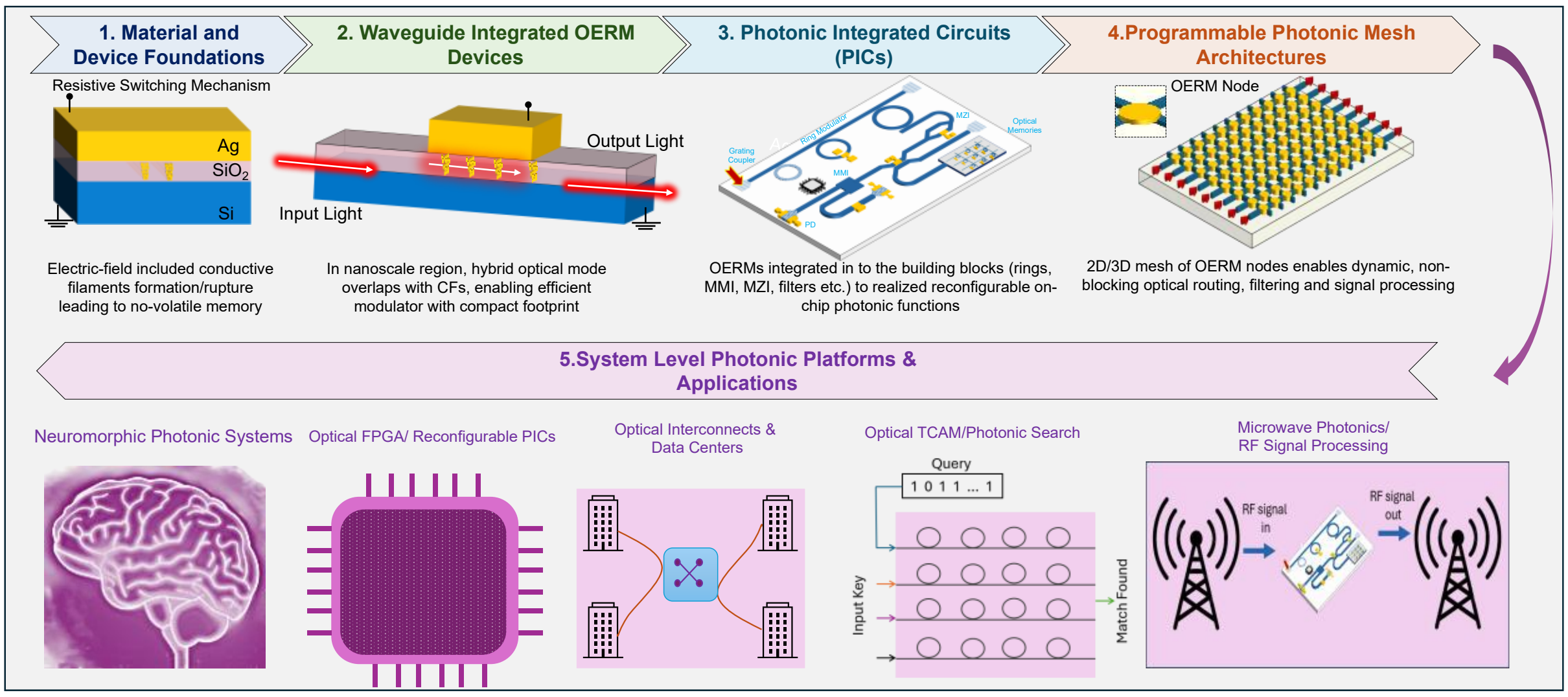

**Fig. 12** Roadmap illustrating the evolution of OERMs from material-level switching mechanisms to system-level programmable photonic platforms. (1) At the material and device level, nanoscale Ag-$SiO_2$-Si structures exhibit resistive switching that enables non-volatile modulation of optical properties. (2) Integration of OERMs within waveguide structures enhances light–matter interaction, enabling efficient and compact optical modulation. (3) These devices are incorporated into PICs alongside passive and active components to realize reconfigurable on-chip functionalities. (4) Scaling to programmable photonic mesh architectures allows dynamic, non-volatile, and non-blocking control of optical routing, filtering, and signal processing through dense networks of OERM nodes. (5) At the system level, these capabilities enable advanced applications such as optical field-programmable gate arrays, neuromorphic photonic computing, reconfigurable optical interconnects, OTCAM/photonic search, microwave photonics/RF signal processing, optical interconnects/data center highlighting the pathway toward intelligent, energy-efficient, and scalable photonic technologies.

## 5. System-Level Functions and Comparison with Competing Non-Volatile Photonic Memories

The development of OERMs has evolved from fundamental material studies to device-level demonstrations and, more recently, toward their integration within PICs. Beyond individual switching elements, the true impact of OERMs lies in their ability to enable system-level functionalities, where memory and optical signal processing are co-integrated within a unified platform. This capability introduces a new paradigm of non-volatile, programmable photonics, enabling adaptive and energy-efficient operation across a wide range of applications.

***Non-Volatile Optical Memory: -*** OERMs provide a promising route toward compact, energy-efficient, and programmable non-volatile optical memory systems by combining electrically controlled resistive switching with optical readout functionality. Unlike conventional electro-optic devices that require continuous biasing, OERMs retain their programmed optical state without static power consumption, making them highly attractive for large-scale programmable PICs. Early demonstrations of integrated photonic memory based on phase-change materials (PCMs), particularly the pioneering work by Carlos Ríos (Nature Photonics, 2015), established the feasibility of non-volatile optical storage and reconfigurable photonic functionality[11]. Subsequent studies by Johannes Feldmann (Nature, 2019; 2021) further extended photonic memory concepts toward photonic computing and neuromorphic architectures[12,18]. More recently, OERM-based platforms have enabled low-voltage, nanoscale, and highly scalable optical memory technologies through diverse material and device approaches. Representative demonstrations include III-V/Si memresonator arrays for integrated optical memory and parallel photonic processing[80], all-silicon avalanche-trapping memories employing microring resonators[132], graphene/$SiO_2$ memristive electroluminescent memory devices[133], and thermally engineered electro-optic memristors utilizing phase-change materials[13]. These architectures demonstrate non-volatile optical state retention, wavelength programmability, multilevel switching capability, and energy-efficient optical modulation.

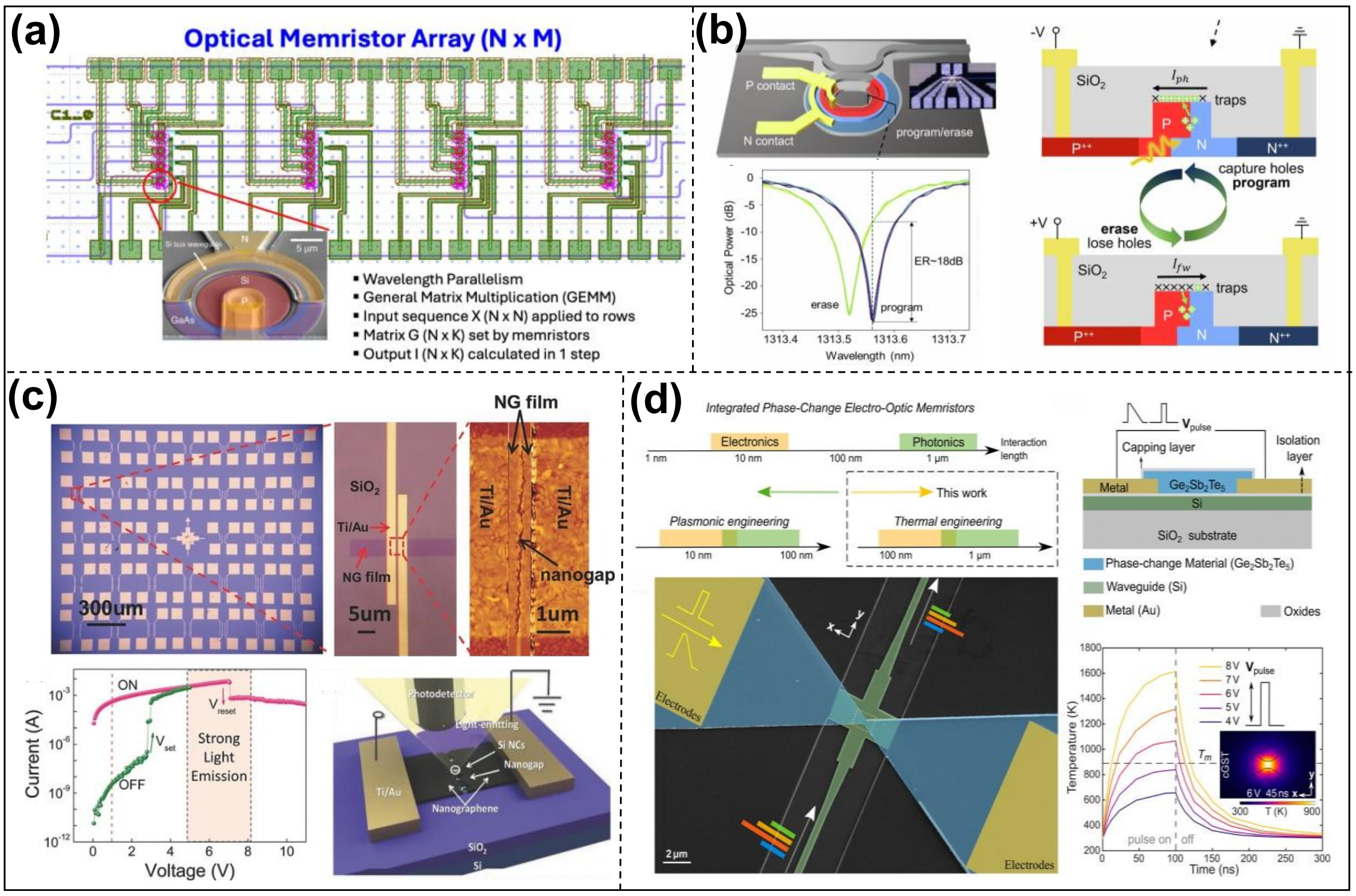

**Fig. 13. Representative demonstrations of non-volatile optical memory (a)** N × M optical memristor array based on III-V/Si mem-resonators, illustrating scalable photonic memory architectures for parallel optical signal processing and integrated memory–computation functionality. **(b)** All-silicon non-volatile optical memory based on photon-avalanche-induced charge trapping in a microring resonator structure, showing device configuration, Si-SiO2 interface traps, program/erase mechanisms, temperature-dependent resonant wavelength switching, optical transmission spectra, and dark-current characteristics associated with avalanche-induced memory operation. *Adapted from [[132]]* **(c)** Planar graphene/$SiO_2$ memristive optical memory and electroluminescent device architecture, including optical and AFM images of the graphene nanogap memristor, resistive switching characteristics, electroluminescence behavior, and the corresponding electrical–optical measurement configuration. *Adapted from [[133]] with permission form John Wiley and Sons* **(d)** Energy-efficient integrated electro-optic memristor employing thermally engineered phase-change materials, showing the device concept, fabricated electro-optic memristor structure, thermal switching dynamics, simulated temperature evolution within the GST active region, and phase-transition-assisted non-volatile optical memory operation. Collectively, these demonstrations highlight the evolution of non-volatile optical memories toward scalable, energy-efficient, multifunctional, and programmable photonic memory systems for integrated photonics, neuromorphic computing, and in-memory optical processing applications. *Adapted from [[13]], licensed under CC BY 4.0.*

Figure 13(a) illustrates a scalable N × M optical memristor array architecture based on III–V/Si memresonators, highlighting the integration of non-volatile photonic memory elements with parallel optical signal-processing functionality. The array architecture demonstrates wavelength-parallel operation and matrix-based optical computation, representing an important step toward photonic in-memory computing and large-scale programmable photonic systems. Figure 13(b) presents an all-silicon non-volatile optical memory based on photon-avalanche-induced charge trapping in a microring resonator. In this device, avalanche-generated carriers become trapped at the Si-$SiO_2$ interface, thereby modifying the effective refractive index and inducing persistent resonance shifts. The demonstrated program/erase operation, wavelength hysteresis, and temperature-dependent switching behaviour provide a CMOS-compatible approach toward fully silicon-integrated optical memory platforms[132]. Figure 13(c) shows a planar graphene/$SiO_2$ memristive optical memory and electroluminescent device based on nanoscale graphene nanogaps. The device combines resistive switching and electrically tunable light emission within a compact planar architecture. The measured I-V hysteresis and electroluminescence characteristics reveal the coexistence of memory and optical functionality, highlighting the potential of graphene-based memristive systems for multifunctional optoelectronic memory applications[133]. Figure 13(d) demonstrates an energy-efficient integrated electro-optic memristor employing thermally engineered phase-change materials. By combining plasmonic and thermal engineering, the device achieves localized phase-transition switching with significantly reduced operating energy. The corresponding thermal simulations and optical modulation characteristics illustrate efficient non-volatile electro-optic memory operation suitable for dense integrated photonic systems[13]. Compared with conventional PCM-based photonic memories, OERMs generally offer lower switching voltage, reduced operating energy, improved CMOS compatibility, and enhanced scalability toward dense integrated photonic systems[91,134]. Furthermore, their multilevel memory capability enables high-density optical storage and analog photonic memory operation, making them highly suitable for optical buffering, programmable photonic processors, configuration storage, neuromorphic photonics, and in-memory optical computing applications[18,71,87].

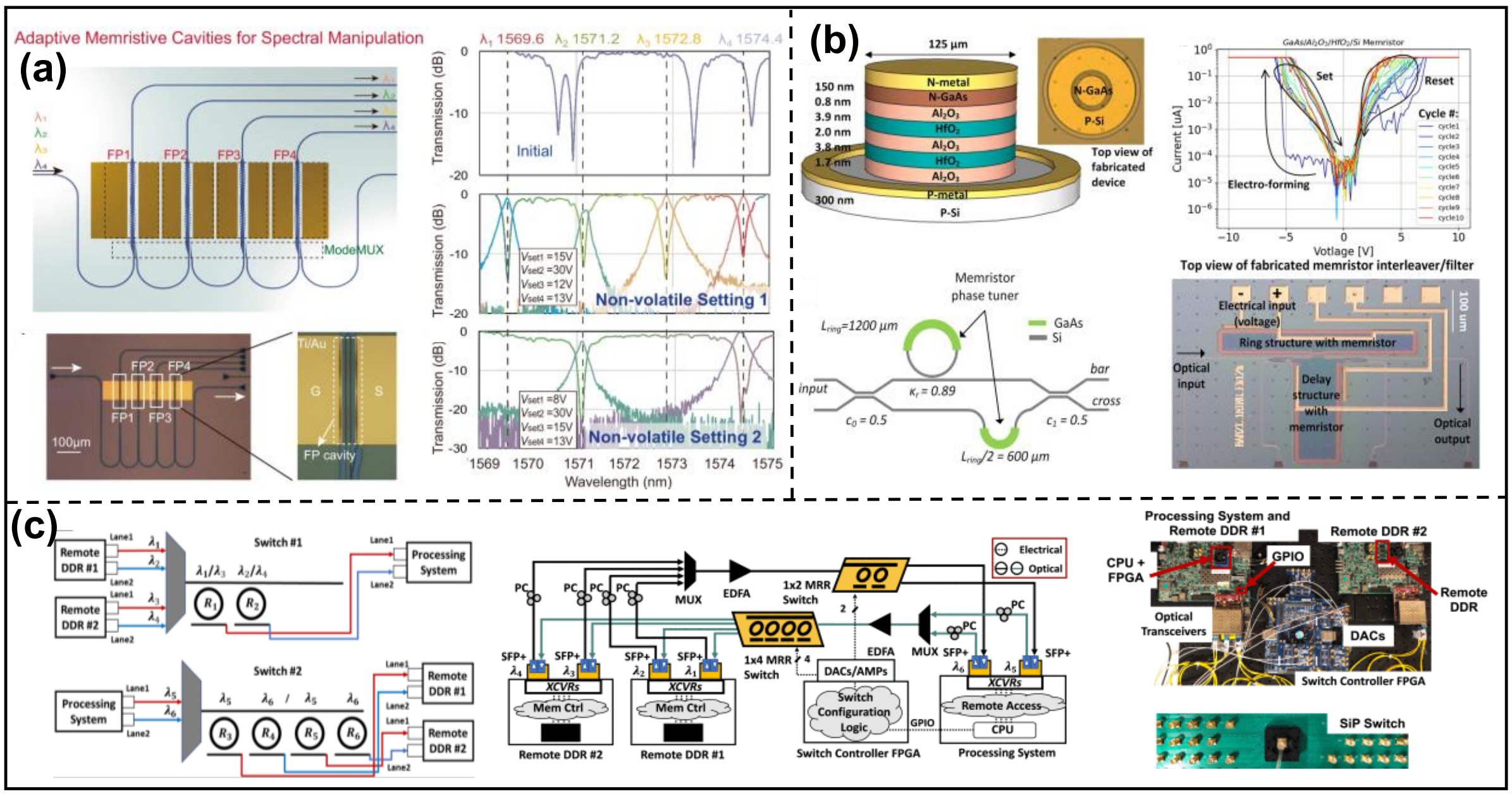

**Fig. 14. Reconfigurable switching and routing networks enabled by non-volatile photonic memory technologies.** (a) PZT optical memristor-based programmable photonic switching architectures demonstrating dynamically reconfigurable 4 × 4 optical routing, adaptive spectral manipulation, and cascaded Fabry-Pérot cavity control through electrically programmed non-volatile phase states. *Adapted from [96], licensed under CC BY 4.0* (b) Heterogeneous III-V/Si memristive photonic circuits employing embedded $HfO_2/Al_2O_3$ semiconductor–insulator–semiconductor capacitor (SISCAP) structures for non-volatile optical switching and programmable ring-assisted Mach-Zehnder interferometer (RAMZI) filtering with memristive phase tuning and zero static power consumption. *Adapted from [78]. (c) Photonic switched optically connected memory architecture illustrating dynamically reconfigurable silicon photonic switching for remote DDR memory access, high-bandwidth optical interconnects, and scalable photonic memory-network integration for data-intensive computing systems. Adapted from [135].*

***Reconfigurable Switching and Routing Networks: -*** The integration of OERMs into waveguide-based and interferometric photonic circuits enables dynamically programmable switching, routing, and spectral reconfiguration functionalities for large-scale photonic integrated systems. By electrically controlling the resistive and optical transmission states of individual OERM elements, optical signals can be selectively routed, filtered, and redistributed through complex photonic network architectures. In contrast to conventional thermo-optic and free-carrier-based switching mechanisms that require continuous bias power to maintain their operational state, OERMs provide non-volatile state retention, allowing switching configurations and routing pathways to remain stable without static power consumption[87,91,134]. Recent advances in programmable photonic platforms, including PZT optical memristor-based MZI switching networks, as depicted in Fig. 14(a), have demonstrated scalable optical routing and adaptive spectral manipulation through electrically programmed non-volatile phase states[96]. Similarly, heterogeneous III-V/Si memristive ring-assisted Mach-Zehnder interferometer filters and programmable switching architectures shown in Fig. 14(b) demonstrate energy-efficient non-volatile optical routing and filtering with zero static power consumption[78]. Furthermore, silicon photonic switched optically connected memory systems illustrated in Fig. 14(c) highlight the potential of dynamically reconfigurable photonic interconnects for remote memory access, optical packet routing, and high-bandwidth data-center communication networks[135]. These developments collectively demonstrate the growing importance of OERMs in enabling scalable, low-power, and programmable photonic routing networks for next-generation optical interconnects, in-memory photonic computing, neuromorphic photonics, and intelligent PICs.

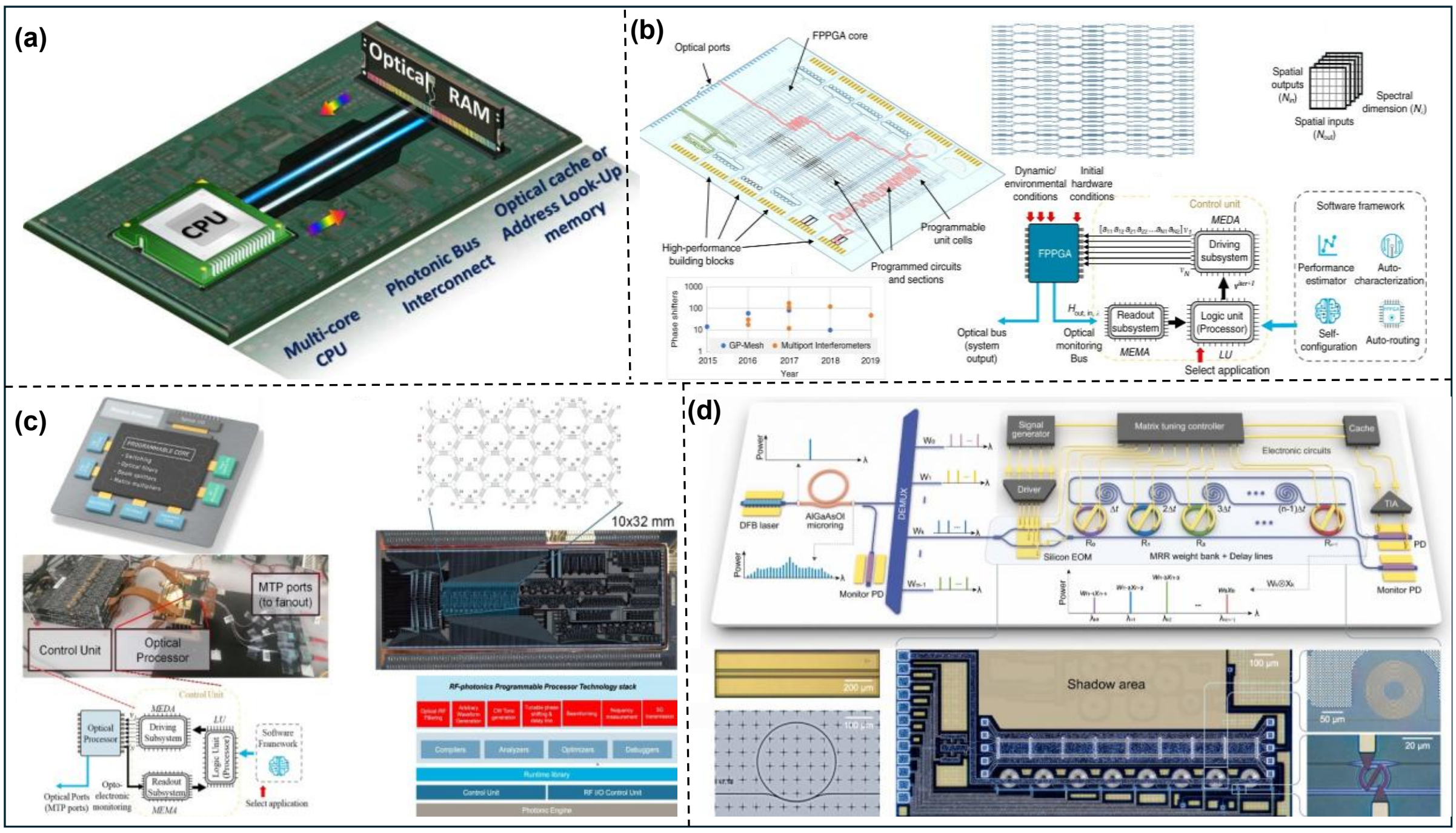

**Fig. 15. Representative programmable photonic processor architectures and field-programmable photonic integrated circuit (FPPGA) platforms for large-scale reconfigurable optical computing and signal processing.** (a) Conceptual architecture of an all-optical RAM cell employing semiconductor optical amplifier (SOA) MZI access gates, AWGs, and optical flip-flop memory units for high-speed optical data storage and random-access operation. The architecture demonstrates wavelength-assisted optical memory functionality for ultrafast photonic buffering, optical interconnects, and programmable photonic computing applications. *Adapted from Fraunhofer HHI technology demonstration reported in PIC Magazine [*[136]*]* (2019). (b) Multipurpose self-configurable programmable photonic circuit employing a field-programmable photonic gate array (FPPGA) architecture with a hexagonal waveguide mesh core, integrated phase shifters, electronic control subsystem, and software-defined optical routing for dynamically reconfigurable photonic signal processing and multifunctional optical computation. *Adapted from [*[14]*], licensed under CC BY 4.0* (c) General-purpose programmable photonic processor architecture for advanced radiofrequency and microwave photonic applications, illustrating the reconfigurable waveguide mesh core, optical input/output interfaces, integrated control electronics, and software-defined photonic processing framework enabling scalable programmable photonic computation. *Adapted from [*[16]*], licensed under CC BY 4.0* (d) Microcomb-based integrated photonic processing unit (PPU) incorporating monolithically integrated electro-optic modulators, microring resonator weight banks, photonic delay lines, and microcomb sources for high-density parallel optical processing and integrated photonic computing applications. *Adapted from [*[17]*], licensed under CC BY 4.0*. Collectively, these architectures demonstrate the evolution of programmable photonic processors toward scalable, software-defined, and energy-efficient optical computing platforms for optical FPGA systems, microwave photonics, neuromorphic computing, and next-generation photonic information processing.

***Programmable Photonic Processors (Optical FPGA): -*** Programmable photonic processors represent an important evolution of OERM-enabled photonic integrated circuits, where non-volatile photonic memory elements are incorporated within mesh-based and interferometric architectures to realize dynamically reconfigurable and software-defined photonic systems. By electrically programming the resistive and optical states of OERM elements, these architectures enable configurable optical routing, filtering, wavelength multiplexing, phase control, and optical logic operations without requiring continuous electrical bias[85,87,137,138].

The non-volatile nature of OERMs significantly reduces static power consumption and enables persistent photonic state retention, which is particularly advantageous for large-scale programmable photonic integrated circuits and optical FPGA platforms[91,139]. Figure 15(a) illustrates a conceptual all-optical RAM architecture employing semiconductor optical amplifier (SOA)-based MZI access gates, AWGs, and optical flip-flop memory units for ultrafast optical data storage and random-access functionality. The architecture demonstrates wavelength-assisted optical memory operation and photonic bus interconnectivity suitable for optical buffering and programmable photonic computing systems[136]. Figure 15(b) presents a multipurpose self-configurable programmable photonic circuit based on a field-programmable photonic gate array (FPPGA) employing a large-scale hexagonal waveguide mesh. Integrated phase shifters, electronic control circuits, and software-defined routing algorithms enable dynamic self-configuration and multifunctional optical signal processing within a unified programmable photonic platform[14]. Figure 15(c) demonstrates a general-purpose programmable photonic processor architecture designed for advanced microwave photonics and radiofrequency signal-processing applications. The architecture integrates a scalable waveguide mesh core, optical input/output interfaces, programmable control electronics, and software-defined photonic processing frameworks, enabling dynamically reconfigurable optical computation and broadband RF signal manipulation[16]. Figure 15(d) illustrates a microcomb-based integrated photonic processing unit (PPU) incorporating monolithically integrated electro-optic modulators, microring resonator weight banks, microcomb sources, and photonic delay lines for massively parallel optical computation. The integrated architecture enables high-density matrix operations and programmable optical signal processing for photonic AI accelerators and in-memory optical computing applications[17]. Collectively, these architectures demonstrate the transition from fixed-function photonic integrated circuits toward scalable, software-defined, and energy-efficient programmable photonic processors, where OERMs can serve as key enabling elements for optical FPGA systems, neuromorphic photonics, microwave photonics, and intelligent photonic computing platforms.

***Neuromorphic and In-Memory Photonic Computing: -*** OERMs can emulate synaptic functionality by providing multi-level and analog conductance states that dynamically modulate optical transmission, absorption, or phase. When integrated within interferometric meshes such as MZI networks or resonant photonic architectures, these programmable states serve as optical weights for matrix-vector multiplication (MVM), which forms the fundamental computational operation in artificial neural networks[86,140,141]. Unlike conventional volatile tuning approaches, the synaptic weights in OERM-based systems are stored in electrically programmed resistive states and accessed optically, enabling non-volatile in-memory photonic computation without continuous electrical bias[107]. Figure 16(a) illustrates a coherent optical neural network (ONN) based on programmable nanophotonic circuits employing cascaded MZI meshes and optical interference units (OIUs) for matrix multiplication and deep-learning inference. The architecture demonstrates programmable SU (4) optical transformations, integrated phase-shifter control, and optical interference-based neural computation, establishing one of the earliest large-scale demonstrations of photonic deep-learning acceleration using coherent integrated photonics[15]. Figure 16(b) presents an electrically programmable in-memory photonic–electronic dot-product engine employing non-volatile GST-based photonic weight banks for scalar multiplication and matrix operations. The architecture combines integrated photonic circuits with electrically programmable phase-change memory cells, enabling direct optical computation using locally stored synaptic weights. The demonstrated reversible programming dynamics, scalar multiplication accuracy, and low computational error distributions highlight the feasibility of photonic in-memory computing systems with integrated non-volatile memory functionality[31]. Figure 16(c) demonstrates an ultra-compact multi-task in-memory optical processor based on a shared optical core (SOC) architecture. The system integrates in-memory optical computation with numerical modelling frameworks based on Rayleigh-Sommerfeld propagation matrices and complex-valued neural networks. Experimental demonstrations, SEM images, and packaged-chip implementations illustrate compact multifunctional optical computing with efficient photonic hardware utilization[142]. Figure 16(d) shows a single-chip photonic deep neural network with forward-only training implemented on a large-scale photonic integrated circuit. The architecture incorporates coherent matrix–vector units (CMXUs), integrated coherent receivers, programmable optical distribution networks, and MZI-mesh-based optical computation blocks for large-scale neural-network inference and training acceleration. The demonstrated system highlights the growing maturity of integrated photonic AI accelerators based on programmable photonic hardware[143]. Compared with thermo-optic and phase-change-material-based approaches, OERMs provide several unique advantages, including low-voltage electrical programmability, compact device footprint, non-volatile memory retention, and compatibility with high-density photonic integration. From a system-level perspective, OERM-enabled photonic computing architectures offer several important benefits:

- In-memory MVM through locally stored synaptic weights, minimizing data transfer overhead
- Analog computing capability with continuously tunable weight states for improved computational efficiency
- Event-driven operation compatible with spiking and neuromorphic photonic systems

• Significant energy savings due to zero static holding power for programmed states

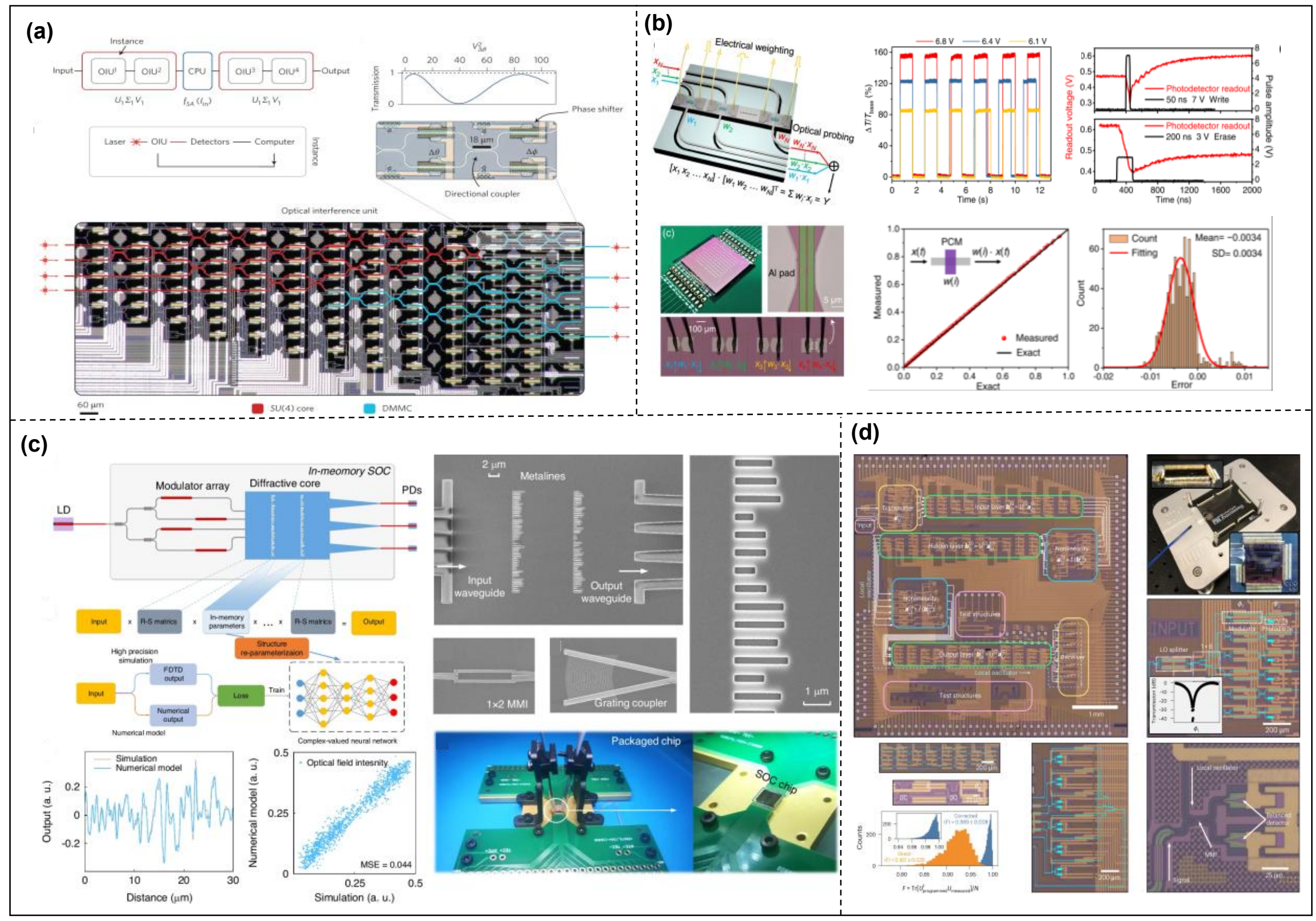

**Fig. 16. Representative demonstrations of neuromorphic and in-memory photonic computing architectures.** **(a)** Optical neural network (ONN) based on coherent nanophotonic circuits demonstrating matrix multiplication and programmable optical interference units (OIUs) for deep learning applications. The figure includes the schematic of a two-layer ONN, experimental feedback and control setup, optical micrograph of the programmable nanophotonic processor implementing SU (4) rotations using MZIs, and phase-shifter tuning characteristics. *Adapted from [15], permission from springer nature.* **(b)** Electrically programmable in-memory photonic–electronic dot-product engine employing non-volatile GST-based photonic weight banks for scalar multiplication and matrix operations. The figure illustrates the computing architecture, fabricated silicon photonic chip, reversible programming characteristics of GST memory cells, temporal switching dynamics, scalar multiplication performance, and computational error distribution. *Adapted from [31], licensed under CC BY 4.0* **(c)** Ultra-compact multi-task in-memory optical processor based on a SOC. The figure presents the on-chip architecture, numerical modelling framework using Rayleigh-Sommerfeld propagation matrices and complex-valued neural networks, comparison between numerical and FDTD simulations, SEM images of fabricated components, and the packaged SOC chip for optoelectronic operation. *Adapted from [142], licensed under CC BY 4.0* **(d)** Single-chip photonic deep neural network with forward-only training implemented on PIC. The figure shows the fabricated PIC and photonic packaging, optical signal and local oscillator (LO) distribution network, MZI-mesh-based coherent matrix–vector unit (CMXU), experimentally measured unitary fidelity distributions, and integrated coherent receiver (ICR) architecture for optical neural computation. *Adapted from [143], permission from springer nature.*

Despite these advantages, several challenges remain for practical large-scale deployment. Device-to-device variability originating from stochastic filament formation dynamics can affect weight precision and repeatability. In addition, write noise, temporal drift, and the trade-off between optical loss and modulation depth can degrade signal fidelity and neural network accuracy. Recent strategies to address these limitations include write-verify programming methods, differential weight encoding schemes, redundant memory architectures, and circuit-level calibration and feedback loops[90,134]. Overall, OERM-based photonic hardware represents a promising pathway toward scalable, energy-efficient, and non-volatile photonic AI accelerators capable of combining optical processing speed with integrated memory functionality for next-generation neuromorphic and in-memory photonic computing systems.

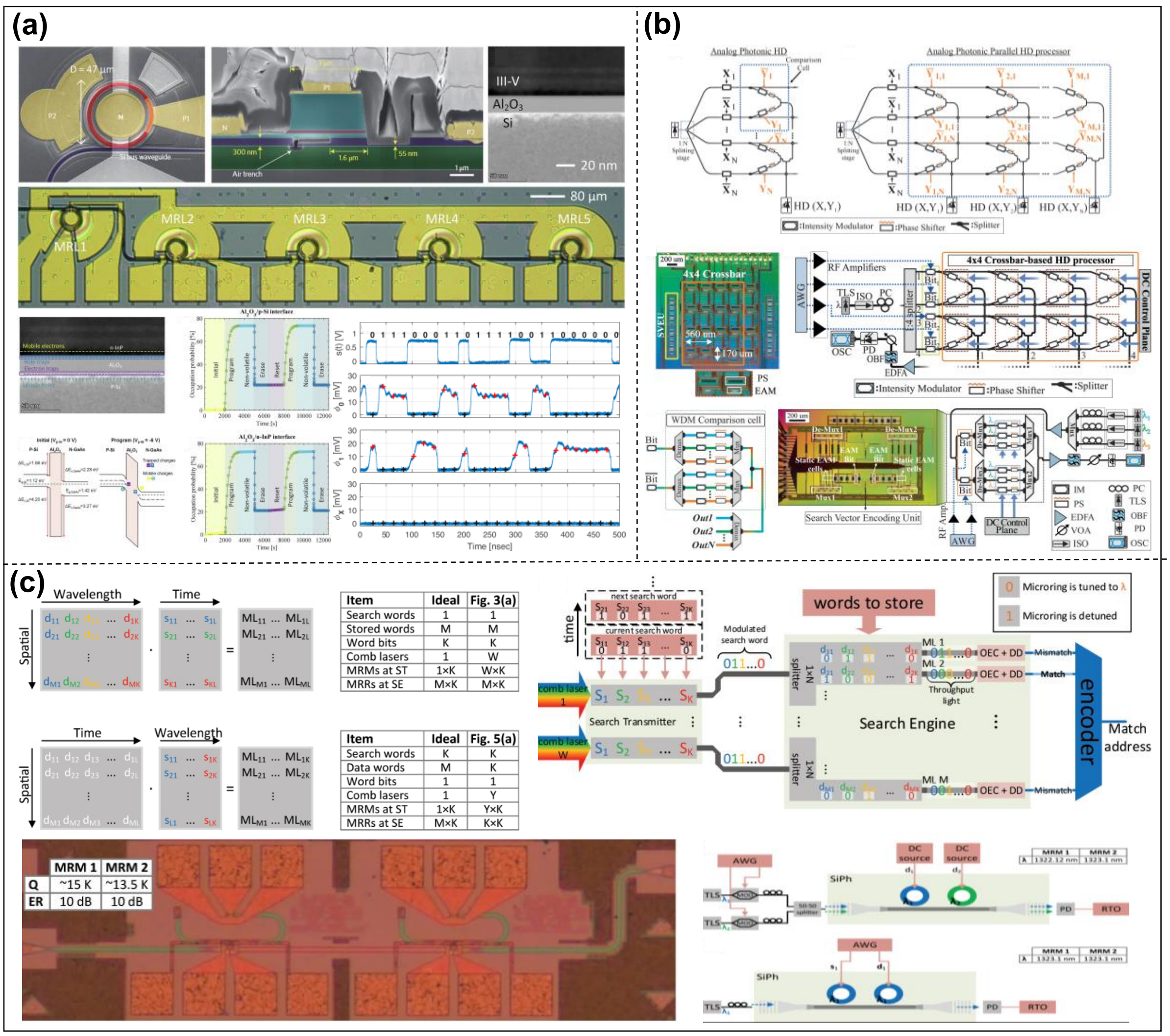

**Fig. 17. Representative architectures and system-level implementations of optical ternary content-addressable memories (O-TCAMs) and programmable photonic search engines enabled by non-volatile photonic memory technologies. (a)** Heterogeneous III-V/Si micro-ring laser arrays integrated with multi-state non-volatile optical memory for programmable O-TCAM operation, showing electrically programmable wavelength control, multi-level optical memory states, and scalable photonic memory arrays for ultrafast optical search and in-memory photonic computing. *Adapted from [144]* **(b)** High-speed nonlinear optical vector processing and photonic content-matching architectures based on linear silicon photonic crossbar circuits, illustrating optical memory operations, string similarity functions, and 50 Gb/s photonic vector processing for neuromorphic and associative photonic computing applications. *Adapted from [145], licensed under CC BY 4.0* **(c)** WDM and TDM O-TCAM architectures employing microring-resonator-based silicon photonic circuits for scalable optical search, multiplexed content-addressable memory functionality, and programmable photonic computing. Adapted from []. *Adapted from [146], licensed under CC BY 4.0.*

***Optical Ternary Content Addressable Memories (O-TCAMs): -*** O-TCAMs represent one of the most promising system-level applications of OERMs and programmable photonic circuits, enabling ultrafast parallel search and associative computing directly in the optical domain. Unlike conventional electronic CAMs, which suffer from bandwidth limitations and high-power consumption due to repeated optical-to-electrical conversions, photonic TCAMs exploit wavelength, spatial, and temporal multiplexing to perform high-speed content matching with significantly enhanced throughput[144,146]. Recent demonstrations based on silicon photonic crossbar architectures, microring resonators, electro-absorption modulators, and non-volatile photonic memories have shown the feasibility of scalable O-TCAM platforms for next-generation optical computing and data-center applications. Figure 17(a) presents a heterogeneous III-V/Si micro-ring laser array integrated with multi-state non-volatile optical memory for programmable O-TCAM functionality. The architecture combines electrically programmable memristive photonic memory with wavelength-tunable micro-ring laser arrays, enabling scalable associative search operations, multi-level optical memory states, and high-density photonic search functionality. The demonstrated platform achieved ~40 dB extinction ratio, 1024 programmable optical states, and femtojoule-per-symbol energy consumption while eliminating static tuning power, highlighting the strong potential of OERM-enabled photonic memories for scalable optical TCAM systems[144]. Figure 17(b) illustrates nonlinear optical vector processing and photonic content-matching architectures based on linear silicon photonic crossbar circuits for high-speed memory and string similarity operations. The demonstrated system employs programmable optical weighting, interferometric crossbar processing, and high-speed optical modulation to realize vector similarity search and Hamming-distance computations at 50 Gb/s data rates. Such architectures demonstrate the capability of photonic associative processors for neuromorphic photonics, AI acceleration, and high-throughput optical search engines[145]. Figure 17(c) shows wavelength-division multiplexed (WDM) and time-division multiplexed (TDM) O-TCAM architectures employing silicon photonic microring-resonator circuits for scalable optical search and associative memory functionality. By exploiting multiplexing in wavelength and temporal domains, the architecture enables parallel optical comparison operations with enhanced search density and throughput. Experimental demonstrations confirmed scalable O-TCAM operation with programmable photonic memory functionality and multiplexed photonic search-engine implementation[146]. Collectively, these developments establish O-TCAMs as a key pathway toward energy-efficient in-memory photonic computing, ultrafast optical packet processing, AI accelerators, and scalable programmable photonic systems. Compared with conventional electronic TCAMs, O-TCAM architectures enabled by OERMs offer several important advantages, including reduced latency, massively parallel search operations, lower static power consumption through non-volatile state retention, and compatibility with wavelength-multiplexed photonic interconnects. These attributes make O-TCAMs highly attractive for future optical networking, neuromorphic information processing, edge-AI accelerators, and photonic database-search applications.

**Comparison with Competing Non-Volatile Photonic Technologies**

Non-volatile functionality in integrated photonics has been pursued through several material platforms, most prominently PCMs, ferroelectric thin films, magneto-optic systems, and, more recently, memristive approaches such as OERMs. Although these technologies share the objective of retaining an optical state without continuous power, they rely on fundamentally different physical mechanisms, which leads to distinct trade-offs in switching energy, speed, endurance, optical loss, and integration scalability[9,20,87,134]. The comparison of various non-volatile photonic memory technologies is illustrated in Fig. 18.

Phase-change photonic memories based on chalcogenides such as GST represent the most mature approach. The large refractive index contrast between amorphous and crystalline phases enables high extinction ratios and reliable multilevel operation in integrated photonic structures. Demonstrations of non-volatile optical memory and computing, including the seminal work by Carlos Ríos and subsequent photonic neural network implementations by Johannes Feldmann, have established PCM as a benchmark platform[8,11,12,18,107]. However, the reliance on thermally induced phase transitions imposes a significant energy overhead, typically in the picojoule to nanojoule range, and introduces thermal crosstalk and material fatigue that can limit endurance[8,11,107].

Ferroelectric photonic devices, particularly those based on $HfO_2$-derived thin films, achieve non-volatility through polarization switching[96,147–149]. These systems exhibit fast response times, potentially reaching the picosecond regime, and comparatively low switching energy. Nevertheless, the achievable optical modulation depth is often limited, and large-scale integration with silicon photonics remains less mature than PCM-based approaches[96,148].

Magneto-optic devices offer intrinsically non-volatile operation via magnetic domain control, but their practical deployment in integrated photonics is constrained by weak magneto-optic interactions at the nanoscale and the need for bulky materials or external magnetic biasing On-chip optical isolation in monolithically integrated non-reciprocal optical resonators[150,151]. As a result, their footprint and integration complexity hinder scalability in dense photonic circuits.

Electro-optic platforms provide ultrafast modulation but are fundamentally volatile, requiring continuous bias to maintain a given state. While hybrid approaches have attempted quasi-non-volatile behavior, they do not offer true state retention without power[152,153]. In this landscape, OERMs introduce a distinct operating paradigm based on electrically controlled resistive switching, where a conductive filament or defect-mediated path defines the stored state[9,20,67,81,87,134,154]. The optical response is modulated by this stored resistive configuration, enabling non-volatile resistive state retention with optical readout. Compared with PCM devices, OERMs can achieve significantly lower switching voltages (typically sub-1–3 V) and reduced energy consumption due to the absence of bulk thermal transitions. Furthermore, their nanoscale active region enables compact device footprints and facilitates integration within slot and hybrid plasmonic waveguides[52,92,101,155]. However, OERMs are not without challenges. The stochastic nature of filament formation introduces variability, which can affect device uniformity and multilevel precision[9,20,90]. Endurance, while improving, remains dependent on material engineering, and optical loss must be carefully managed in highly confined geometries. Despite these limitations, OERMs occupy a favorable position in the trade-off space, offering a balance between energy efficiency, compactness, and functional versatility. The quantitative comparison highlights that PCM devices dominate in modulation strength but at high energy cost, whereas ferroelectric and electro-optic systems excel in speed but lack strong optical contrast or true non-volatility. OERMs, in contrast, offer a balanced combination of low energy, moderate-to-high extinction ratio, and compact footprint, positioning them as a promising platform for energy-efficient programmable photonics[84,90,156].

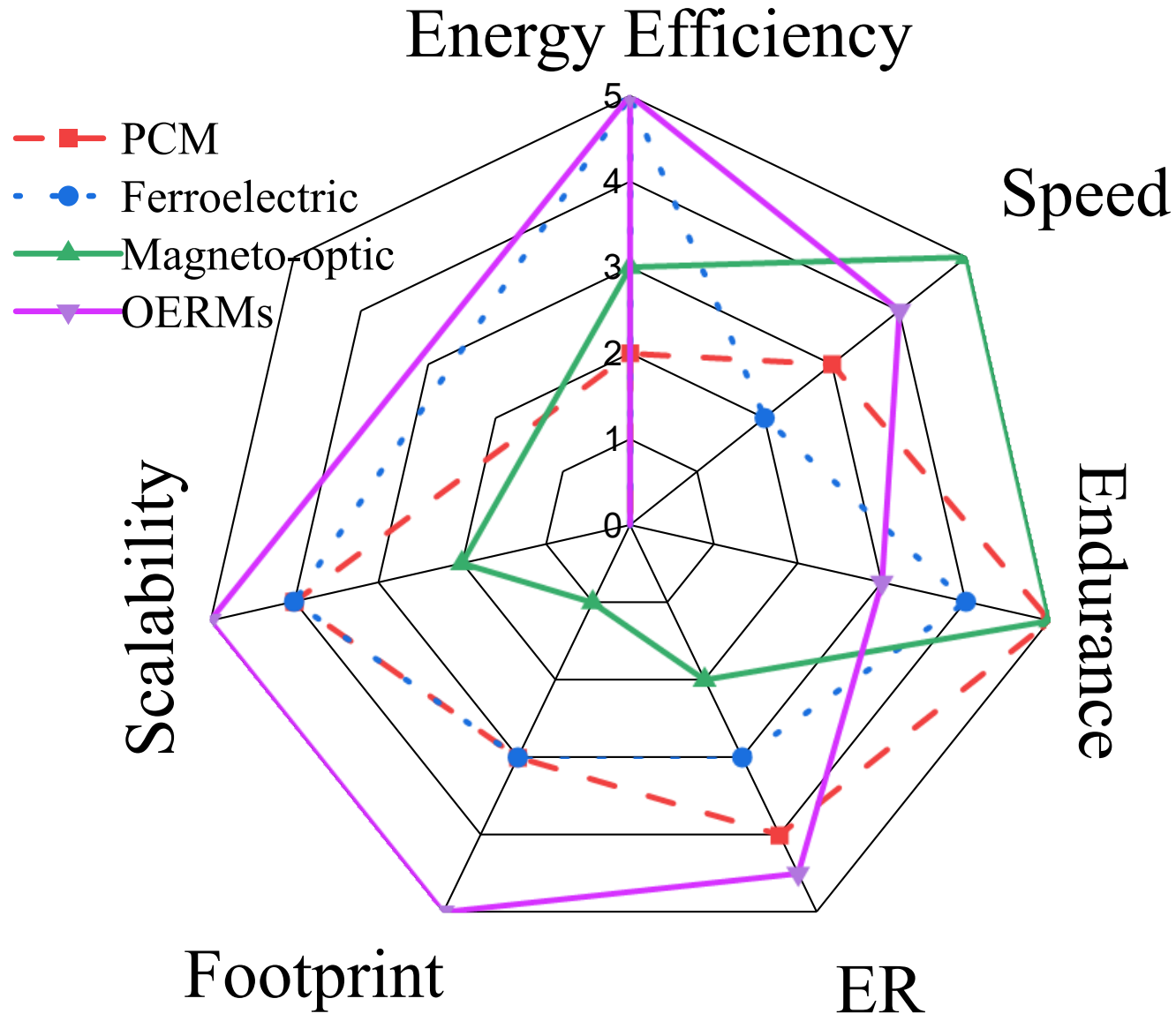

**Fig. 18.** Comparing non-volatile photonic memory technologies, including phase-change-material (PCM), ferroelectric, magneto-optic, and OERM platforms, based on normalized performance metrics including energy efficiency, switching speed, endurance, extinction ratio (ER), footprint, and scalability. Higher normalized values indicate superior performance for the corresponding metric.

## 6. Applications, Challenges, and Future Outlook

The emergence of OERMs has opened new opportunities for integrating non-volatile functionality directly within photonic systems. By combining electrically programmable resistive states with optical readout, OERMs bridge the gap between electronic memory and photonic signal processing, enabling a new class of adaptive and energy-efficient photonic technologies[9,20]. Their unique ability to retain state without continuous power, coupled with compact device footprints and compatibility with integrated photonic platforms, positions them as promising building blocks for next-generation optical systems[91,92,152–154].

### Applications:-

One of the most immediate applications of OERMs lies in on-chip optical memory and data storage, where non-volatile state retention enables persistent configuration of photonic circuits[12,18,80,132,139]. This capability is particularly important for large-scale photonic integrated circuits, where static power consumption associated with conventional tuning mechanisms becomes prohibitive. The multilevel switching characteristics of OERMs further enable high-density optical storage and analog information encoding, which are essential for advanced signal processing and memory-intensive applications[11,18,34]. Beyond memory, OERMs play a critical role in programmable photonic circuits, where they enable dynamic reconfiguration of optical pathways[14,16,67]. In such systems, OERMs can be integrated within waveguide meshes or interferometric networks to realize optical FPGAs, allowing real-time adaptation of circuit functionality[17,78,96]. This programmability is essential for applications in optical communication, signal routing, adaptive filtering, and microwave photonics, where system requirements may change dynamically.

Another rapidly emerging application domain is neuromorphic and in-memory photonic computing, where OERMs can emulate synaptic behavior through multilevel and analog switching[15,31,88,139]. By enabling co-localized storage and computation, OERMs reduce data movement and energy consumption, addressing key limitations of conventional computing architectures. When combined with the inherent parallelism and bandwidth of photonics, this approach offers a powerful platform for implementing artificial intelligence and machine learning systems[107,142]. Furthermore, the ability of OERMs to provide analog weight programmability and non-volatile synaptic storage makes them highly attractive for photonic tensor processing, optical neural networks, and event-driven neuromorphic architectures.

OERMs also enable multiwavelength and parallel photonic systems, particularly in WDM architectures[144,146]. Their integration into wavelength-selective components allows dynamic control over multiple optical channels, enabling parallel data processing, adaptive spectral filtering, optical routing, and efficient utilization of optical bandwidth[144–146,157]. Such capabilities are critical for next-generation communication networks, optical interconnects, and high-capacity data centers. In addition, OERMs show potential in biosensing and lab-on-chip systems, where programmable optical responses can be used to enhance sensitivity and selectivity. By enabling reconfigurable sensing platforms, OERMs can support adaptive detection mechanisms and real-time calibration, which are valuable for biomedical and environmental monitoring applications[158,159]. The integration of OERMs with resonant photonic biosensors may further enable programmable sensing platforms with tunable spectral response and adaptive functionality.

## Challenges and Future Outlook:-

Despite their promising capabilities, several challenges must be addressed to enable the widespread adoption of OERMs in practical photonic systems. One of the primary challenges is device variability, arising from the stochastic nature of filament formation in resistive switching[9,20,90]. This variability can lead to fluctuations in device performance, particularly in multilevel operation, and poses challenges for large-scale integration and precise analog control[15,90,160]. Another critical issue is endurance and reliability, as repeated switching cycles can degrade device performance over time. Although significant progress has been made through material engineering and device optimization, further improvements are required to achieve the levels of reliability demanded by commercial applications[90,161]. The trade-off between optical confinement and loss represents another key challenge, especially in plasmonic and hybrid plasmonic architectures[39,52,101,162]. While strong confinement enhances light–matter interaction and improves modulation efficiency, it also increases optical absorption, which can degrade overall system performance. Thermal effects and crosstalk in densely integrated circuits further complicated system design, as localized heating can influence neighboring devices and affect stability. Additionally, achieving uniform fabrication and scalability across large photonic integrated circuits remains a significant challenge, particularly when incorporating multiple material systems[14,73,152]. Integration compatibility between CMOS fabrication processes and emerging resistive switching materials will also play a crucial role in determining the scalability and commercial viability of OERM technologies.

Looking forward, OERMs are expected to play a central role in the development of programmable and intelligent photonic systems. Advances in material engineering, device design, and integration techniques are likely to improve performance metrics such as endurance, uniformity, and energy efficiency. The integration of OERMs with emerging material platforms, including two-dimensional materials and advanced oxide systems, may further expand their functionality and scalability[16,134,163]. At the system level, the convergence of OERMs with silicon photonics and heterogeneous integration technologies is expected to enable large-scale, reconfigurable photonic circuits with unprecedented flexibility. These developments will pave the way for optical computing architectures, where data processing, storage, and communication are seamlessly integrated within a single platform. In the longer term, OERMs have the potential to enable adaptive and self-configurable photonic hardware, capable of learning and optimizing performance in real time. Such systems could form the foundation of future technologies in artificial intelligence, quantum photonics, optical interconnects, secure photonic communication, and advanced sensing, where programmability and energy efficiency are paramount[156,164,165]. The continued development of low-loss material platforms, scalable fabrication strategies, and hybrid photonic-electronic integration frameworks is expected to accelerate the transition of OERMs from laboratory-scale demonstrations toward practical deployment in next-generation intelligent photonic systems.

## 7. Conclusion

In this review, we have presented a comprehensive overview of OERMs, covering their underlying physical mechanisms, device architectures, and emerging roles in integrated photonic systems. Starting from material-level insights into resistive switching and light-matter interaction, we discussed how OERMs can be engineered into nanoscale photonic devices capable of non-volatile resistive state retention with optical readout. The integration of these devices within silicon photonics and hybrid platforms enables a new class of compact, energy-efficient, and reconfigurable photonic components. At the device level, OERMs demonstrate promising performance in terms of low switching voltage, high extinction ratio, multilevel operation, and compatibility with subwavelength confinement structures. When extended to circuit and system levels, these characteristics enable a wide range of functionalities, including non-volatile optical memory, programmable photonic circuits, neuromorphic computing, and multiwavelength signal processing. Through a detailed comparison with established non-volatile photonic technologies such as phase-change materials, ferroelectric devices, and magneto-optic systems, OERMs have been positioned as a complementary and potentially disruptive platform that balances energy efficiency, scalability, and functionality.

Despite these advances, several challenges remain, particularly in achieving device uniformity, improving endurance, and minimizing optical loss in highly confined geometries. Addressing these issues will require continued progress in material engineering, device optimization, and large-scale integration strategies. In parallel, the development of robust circuit architectures and system-level design methodologies will be essential to fully harness the capabilities of OERMs in practical applications. Looking ahead, OERMs are poised to play a pivotal role in the evolution of programmable and intelligent photonic systems. Their unique ability to combine non-volatile electrical memory with optical functionality opens new pathways toward adaptive, energy-efficient, and highly integrated photonic hardware. As advances continue across materials, devices, and system architectures, OERMs are expected to contribute significantly to future technologies in optical communication, artificial intelligence, neuromorphic computing, and beyond.

## Acknowledgements

The authors would like to thank Hewlett Packard Enterprise (HPE) for the valuable scientific discussions, technical support, and collaborative research environment that contributed to this review work.

## Author contribution

S. Kumar carried the literature survey, figure preparation, manuscript writing, and organization of the review article. S. Cheung provided major scientific discussions, technical suggestions, and continuous guidance during the preparation of the manuscript. Mukesh Kumar provided overall support throughout the work. E. Shin and B. Tossoun contributed through technical discussions, suggestions, and manuscript improvement.

## Conflict of interest

The authors declare no competing interests.

## References

1 Thomson D, Zilkie A, Bowers JE, Komljenovic T, Reed GT, Vivien L *et al.* Roadmap on silicon photonics. *J Opt (United Kingdom)* 2016; **18**. doi:10.1088/2040-8978/18/7/073003.
2 Soref R. The past, present, and future of silicon photonics. *IEEE J Sel Top Quantum Electron* 2006; **12**: 1678–1687.
3 Miller DAB. Device requirements for optical interconnects to silicon chips. *Proc IEEE* 2009; **97**: 1166–1185.
4 Wulf WA, McKee SA. Hitting the memory wall. *ACM SIGARCH Comput Archit News* 1995; **23**: 20–24.
5 Wong HSP, Lee HY, Yu S, Chen YS, Wu Y, Chen PS *et al.* Metal-oxide RRAM. *Proc IEEE* 2012; **100**: 1951–1970.
6 Waser R, Aono M. Nanoionics-based resistive switching memories. *Nat Mater 2007 611* 2007; **6**: 833–840.
7 Apalkov D, Dieny B, Slaughter JM. Magnetoresistive Random Access Memory. *Proc IEEE* 2016; **104**: 1796–1830.
8 Wuttig M, Yamada N. Phase-change materials for rewriteable data storage. *Nat Mater 2007 611* 2007; **6**: 824–832.
9 Yang JJ, Strukov DB, Stewart DR. Memristive devices for computing. *Nat Nanotechnol 2013 81* 2012; **8**: 13–24.
10 Emboras A, Goykhman I, Desiatov B, Mazurski N, Stern L, Shappir J *et al.* Nanoscale plasmonic memristor with optical readout functionality. *Nano Lett* 2013; **13**: 6151–6155.
11 Rios C, Stegmaier M, Hosseini P, Wang D, Scherer T, Wright CD *et al.* Integrated all-photonic non-volatile multi-level memory. *Nat Photonics 2015 911* 2015; **9**: 725–732.
12 Feldmann J, Youngblood N, Wright CD, Bhaskaran H, Pernice WHP. All-optical spiking neurosynaptic networks with self-learning capabilities. *Nat 2019 5697755* 2019; **569**: 208–214.
13 He Y, Farmakidis N, Aggarwal S, Dong B, Lee JS, Wang M *et al.* Energy-Efficient Integrated Electro-Optic Memristors. *Nano Lett* 2024; **24**: 16325–16332.
14 Pérez-López D, López A, DasMahapatra P, Capmany J. Multipurpose self-configuration of programmable photonic circuits. *Nat Commun 2020 111* 2020; **11**: 6359-.
15 Shen Y, Harris NC, Skirlo S, Prabhu M, Baehr-Jones T, Hochberg M *et al.* Deep learning with coherent nanophotonic circuits. *Nat Photonics* 2017; **11**: 441–446.
16 Pérez-López D, Gutierrez A, Sánchez D, López-Hernández A, Gutierrez M, Sánchez-Gomáriz E *et al.* General-purpose programmable photonic processor for advanced radiofrequency applications. *Nat Commun 2024 151* 2024; **15**: 1563-.
17 Bai B, Yang Q, Shu H, Chang L, Yang F, Shen B *et al.* Microcomb-based integrated photonic processing unit. *Nat Commun 2023 141* 2023; **14**: 66-.
18 Feldmann J, Youngblood N, Karpov M, Gehring H, Li X, Stappers M *et al.* Parallel convolutional processing using an integrated photonic tensor core. *Nat 2020 5897840* 2021; **589**: 52–58.
19 Valov I, Waser R, Jameson JR, Kozicki MN. Electrochemical metallization memories - Fundamentals, applications, prospects. *Nanotechnology* 2011; **22**: 254003.

20 D Ielmini. Resistive switching memories based on metal oxides: mechanisms, reliability and scaling. *Semicond Sci Technol* 2016; **31**.https://iopscience.iop.org/article/10.1088/0268-1242/31/6/063002/meta (accessed 12 Jun2023).
21 Strukov DB, Snider GS, Stewart DR, Williams RS. The missing memristor found. *Nature* 2008; **453**: 80–83.
22 Korneluk A, Stefaniuk T. Multilevel Optical Storage, Dynamic Light Modulation, and Polarization Control in Filamented Memristor System. *Adv Mater* 2025; **37**: 2411186.
23 Portner K, Schmuck M, Lehmann P, Weilenmann C, Haffner C, Ma P *et al.* Analog Nanoscale Electro-Optical Synapses for Neuromorphic Computing Applications. *ACS Nano* 2021; **15**: 14776–14785.
24 Vishwanath SK, Kim J. Resistive switching characteristics of all-solution-based Ag/TiO2/Mo-doped In2O3 devices for non-volatile memory applications. *J Mater Chem C* 2016; **4**: 10967–10972.
25 Singh L, Jain S, Kumar M. Electrically writable silicon nanophotonic resistive memory with inherent stochasticity. *Opt Lett* 2019; **44**: 4020.
26 Sawa A. Resistive switching in transition metal oxides. *Mater Today* 2008; **11**: 28–36.
27 Rozenberg MJ, Sánchez MJ, Weht R, Acha C, Gomez-Marlasca F, Levy P. Mechanism for bipolar resistive switching in transition-metal oxides. *Phys Rev B* 2010; **81**: 115101.
28 Xu Q, Schmidt B, Pradhan S, Lipson M. Micrometre-scale silicon electro-optic modulator. *Nat 2005 4357040* 2005; **435**: 325–327.
29 Rios C, Hosseini P, Wright CD, Bhaskaran H, Pernice WHP. On-chip photonic memory elements employing phase-change materials. *Adv Mater* 2014; **26**: 1372–1377.
30 Meng J, Gui Y, Nouri BM, Ma X, Zhang Y, Popescu CC *et al.* Electrical programmable multilevel nonvolatile photonic random-access memory. *Light Sci Appl 2023 121* 2023; **12**: 1–10.
31 Zhou W, Dong B, Farmakidis N, Li X, Youngblood N, Huang K *et al.* In-memory photonic dot-product engine with electrically programmable weight banks. *Nat Commun 2023 141* 2023; **14**: 2887-.
32 Chen R, Fang Z, Perez C, Miller F, Kumari K, Saxena A *et al.* Non-volatile electrically programmable integrated photonics with a 5-bit operation. *Nat Commun 2023 141* 2023; **14**: 1–10.
33 Ríos C, Du Q, Zhang Y, Popescu CC, Shalaginov MY, Miller P *et al.* Ultra-compact nonvolatile phase shifter based on electrically reprogrammable transparent phase change materials. *PhotoniX 2022 31* 2022; **3**: 1–13.
34 Meng J, Gui Y, Nouri BM, Ma X, Zhang Y, Popescu CC *et al.* Electrical programmable multilevel nonvolatile photonic random-access memory. *Light Sci Appl 2023 121* 2023; **12**: 189-.
35 Wuttig M, Bhaskaran H, Taubner T. Phase-change materials for non-volatile photonic applications. *Nat Photonics* 2017; **11**: 465–476.
36 Koch U, Hoessbacher C, Emboras A, Leuthold J. Optical memristive switches. *J Electroceramics* 2017; **39**: 239–250.
37 Chua LO. Memristor - the missing circuit element. *IEEE Trans Circuit Theory* 1971; **18**: 507–519.
38 Pérez D, Gasulla I, Crudgington L, Thomson DJ, Khokhar AZ, Li K *et al.* Multipurpose silicon photonics signal processor core. *Nat Commun* 2017; **8**. doi:10.1038/s41467-017-00714-1.
39 Alam MZ, Aitchison JS, Mojahedi M. A marriage of convenience: Hybridization of surface plasmon and dielectric waveguide modes. *Laser Photon Rev* 2014; **8**: 394–408.
40 Kumar S, Kumar P, Ranjan R. A Metal-Cap Wedge Shape Hybrid Plasmonic Waveguide for Nano-scale Light Confinement and Long Propagation Range. *Plasmonics* 2021; **17**: 95–110.
41 Singh L, Sharma T, Kumar M. Controlled Hybridization of Plasmonic and Optical Modes for Low-Loss Nano-Scale Optical Confinement with Ultralow Dispersion. *IEEE J Quantum Electron* 2018; **54**. doi:10.1109/JQE.2018.2809461.
42 Kumar S, Rajput S, Kaushik V, Babu P, Dev Mishra R, Ranjan R *et al.* Numerical Analysis of Laterally and Vertically Coupled Hybrid Plasmonic Modes in Silicon Tip. *Plasmonics* 2022. doi:10.1007/S11468-022-01657-0.
43 Sulabh, Singh L, Jain S, Kumar M. Optical Slot Waveguide With Grating-Loaded Cladding of Silicon and Titanium Dioxide for Label-Free Bio-Sensing. *IEEE Sens J* 2019; **19**: 6126–6133.
44 Jain S, Srivastava S, Rajput S, Singh L, Tiwari P, Srivastava AK *et al.* Thermally Stable Optical Filtering Using Silicon-Based Comb-Like Asymmetric Grating for Sensing Applications. *IEEE Sens J* 2020; **20**: 3529–3535.
45 Singh L, Tidke S, Kumar M. Guiding and controlling light at nanoscale in field effect transistor. *Appl Phys B Lasers Opt* 2019; **125**: 1–7.
46 Kumar S, Kumar P, Ranjan R. Triangular shape hybrid metal-insulator-metal plasmonic waveguide for low propagation loss at deep subwavelength. *IEEE Trans Nanotechnol* 2022; **21**: 6–15.
47 Kumar A, Kumar S, Mishra RD, Pandey SK, Babu P, Kumar M. Distributed semiconductor heterojunctions of ZnO–Cu2O for ultraviolet photodetection. *Opt Mater (Amst)* 2024; **157**: 116092.
48 Kumar S, Kumar P, Ranjan R. Mode analysis of algaas based hybrid metal insulator plasmonic waveguide with nanoscale confinement. *2021 Natl Conf Commun NCC 2021* 2021. doi:10.1109/NCC52529.2021.9530139.
49 Alam MZ, Wagner SJ, Mojahedi M, Sun X, Aitchison JS. Experimental demonstration of a hybrid plasmonic transverse electric pass polarizer for a silicon-on-insulator platform. *Opt Lett Vol 37, Issue 23, pp 4814-4816* 2012; **37**: 4814–4816.
50 Alam MZ, Mojahedi M, Aitchison JS. Compact and silicon-on-insulator-compatible hybrid plasmonic TE-pass polarizer. *Opt Lett Vol 37, Issue 1, pp 55-57* 2012; **37**: 55–57.
51 Sorger VJ, Lanzillotti-Kimura ND, Ma RM, Zhang X. Ultra-compact silicon nanophotonic modulator with broadband response. *Nanophotonics* 2012; **1**: 17–22.
52 Alam MZ, De Leon I, Boyd RW. Large optical nonlinearity of indium tin oxide in its epsilon-near-zero region. *Science (80- )* 2016; **352**: 795–797.
53 Chen SH, Young SJ, Yang CC, Liu YH. Resistive switching behavior of perovskite BiFeO3 thin films and their two-bit-per-cell applications. *J Mater Sci Mater Electron* 2025; **36**. doi:10.1007/S10854-025-14436-4.
54 Chavan GT, Kim Y, Khokhar MQ, Hussain SQ, Cho EC, Yi J *et al.* A Brief Review of Transparent Conducting Oxides (TCO): The Influence of Different Deposition Techniques on the Efficiency of Solar Cells. *Nanomaterials* 2023; **13**. doi:10.3390/NANO13071226.
55 Huebner CF, Tsyalkovsky V, Bandera Y, Burdette MK, Shetzline JA, Tonkin C *et al.* Nonvolatile optically-erased

colloidal memristors. *Nanoscale* 2015; **7**: 1270–1279.
56 Minnekhanov AA, Emelyanov A V., Lapkin DA, Nikiruy KE, Shvetsov BS, Nesmelov AA *et al.* Parylene Based Memristive Devices with Multilevel Resistive Switching for Neuromorphic Applications. *Sci Reports 2019 91* 2019; **9**: 1–9.
57 Chen P-Z, Chen C-T, Lee C-J, Lee Y-J, Lee Y-J, Huang C-L *et al.* High-bandwidth nonvolatile optical memory based on MMI-integrated racetrack resonator with embedded ReRAM. *Opt Express, Vol 33, Issue 23, pp 48081-48099* 2025; **33**: 48081–48099.
58 Shekhar S, Bogaerts W, Chrostowski L, Bowers JE, Hochberg M, Soref R *et al.* Roadmapping the next generation of silicon photonics. *Nat Commun 2024 151* 2024; **15**: 1–15.
59 Tossoun B, Liang D, Sheng X, Strachan JP, Beausoleil RG. Heterogeneously Integrated Memristive Laser on Silicon with Non-Volatile Wavelength Tuning. 2024.https://arxiv.org/pdf/2401.13757 (accessed 22 May2026).
60 Ielmini D, Waser R. Resistive Switching: From Fundamentals of Nanoionic Redox Processes to Memristive Device Applications. *Resist Switch from Fundam Nanoionic Redox Process to Memristive Device Appl* 2016; : 1–755.
61 Jeong DS, Thomas R, Katiyar RS, Scott JF, Kohlstedt H, Petraru A *et al.* Emerging memories: Resistive switching mechanisms and current status. *Reports Prog Phys* 2012; **75**. doi:10.1088/0034-4885/75/7/076502.
62 Neuro-inspired Computing Using Resistive Synaptic Devices. *Neuro-inspired Comput Using Resist Synaptic Devices* 2017. doi:10.1007/978-3-319-54313-0.
63 Singh L, Sulabh, Kaushik V, Rajput S, Mishra RD, Kumar M. Light Assisted Electro-Metallization in Resistive Switch with Optical Accessibility. *J Light Technol* 2021; **39**: 5869–5874.
64 Sacchi E, Zanetto F, Martinez AI, SeyedinNavadeh SM, Morichetti F, Melloni A *et al.* Integrated electronic controller for dynamic self-configuration of photonic circuits. *Light Sci Appl 2025 141* 2025; **14**: 348-.
65 Joushaghani A, Jeong J, Paradis S, Alain D, Stewart Aitchison J, S Poon JK *et al.* Wavelength-size hybrid Si-$VO_2$ waveguide electroabsorption optical switches and photodetectors. *Opt Express, Vol 23, Issue 3, pp 3657-3668* 2015; **23**: 3657–3668.
66 Miscuglio M, Sorger VJ. Photonic tensor cores for machine learning. *Appl Phys Rev* 2020; **7**: 31404.
67 Santosh Kumar, Ashutosh Kumar, Dev Mishra R, Kumar Pandey S, Prem Babu, Mukesh Kumar. Reconfigurable multiwavelength nanophotonic circuit based on a low-voltage, optically readable engineered resistive switch. *Nanoscale* 2025. doi:10.1039/D5NR00463B.
68 Zhang H, Zhou L, Lu L, Xu J, Wang N, Hu H *et al.* Miniature Multilevel Optical Memristive Switch Using Phase Change Material. *ACS Photonics* 2019; **6**: 2205–2212.
69 Ríos C, Youngblood N, Cheng Z, Le Gallo M, Pernice WHP, Wright CD *et al.* In-memory computing on a photonic platform. *Sci Adv* 2019; **5**. doi:10.1126/SCIADV.AAU5759;SUBPAGE:STRING:FULL.
70 Gu T, Majumdar A, Teng J. Programmable nano-optics and photonics. *Nanophotonics* 2024; **13**: 2047–2049.
71 Li J, Chen W, Su Y, Guo X. Photonic memristive devices for in-memory matrix operation. *PhotoniX 2026 71* 2026; **7**: 26-.
72 Xia F, Wang H, Xiao D, Dubey M, Ramasubramaniam A. Two-dimensional material nanophotonics. *Nat Photonics 2014 812* 2014; **8**: 899–907.
73 Geim AK, Grigorieva I V. Van der Waals heterostructures. *Nat 2013 4997459* 2013; **499**: 419–425.
74 Bao Q, Loh KP. Graphene Photonics, Plasmonics, and Broadband Optoelectronic Devices. *ACS Nano* 2012; **6**: 3677–3694.
75 Wang QH, Kalantar-Zadeh K, Kis A, Coleman JN, Strano MS. Electronics and optoelectronics of two-dimensional transition metal dichalcogenides. *Nat Nanotechnol 2012 711* 2012; **7**: 699–712.
76 Mak KF, Shan J. Photonics and optoelectronics of 2D semiconductor transition metal dichalcogenides. *Nat Photonics 2016 104* 2016; **10**: 216–226.
77 Babicheva VE, Lavrinenko A V. Plasmonic modulator optimized by patterning of active layer and tuning permittivity. *Opt Commun* 2012; **285**: 5500–5507.
78 Cheung S, Tossoun B, Yuan Y, Peng Y, Hu Y, Sorin W V. *et al.* Energy efficient photonic memory based on electrically programmable embedded III-V/Si memristors: switches and filters. *Commun Eng 2024 31* 2024; **3**: 49-.
79 Cheung S, London Y, Yuan Y, Tossoun B, Peng Y, Hu Y *et al.* Heterogeneous III-V/Si micro-ring laser array with multi-state non-volatile memory for ternary content-addressable memories. *Nat Commun 2025 161* 2025; **16**: 1–11.
80 Tossoun B, Liang D, Cheung S, Fang Z, Sheng X, Strachan JP *et al.* High-speed and energy-efficient non-volatile silicon photonic memory based on heterogeneously integrated memresonator. *Nat Commun 2024 151* 2024; **15**: 1–8.
81 Kumar S, Mishra RD, Kumar A, Babu P, Pandey S, Kumar M. Double-Slot Nanophotonic Platform for Optically Accessible Resistive Switching with High Extinction Ratio and High Endurance. *ACS Photonics* 2023; **10**: 4071–4078.
82 Singh L, Srivastava S, Rajput S, Kaushik V, Mishra RD, Kumar M *et al.* Optical switch with ultra high extinction ratio using electrically controlled metal diffusion. *Opt Lett Vol 46, Issue 11, pp 2626-2629* 2021; **46**: 2626–2629.
83 Dev Mishra R, Kumar Pandey S, Babu P, Kumar S, Kumar A, Mohanta N *et al.* Nanophotonic resistive switch based on tapered copper-silicon structure with low power and high extinction ratio. *Opt Laser Technol* 2024; **175**: 110833.
84 Fang Z, Chen R, Tossoun B, Cheung S, Liang D, Majumdar A. Non-volatile materials for programmable photonics. *APL Mater* 2023; **11**: 100603.
85 Emboras A, Alabastri A, Lehmann P, Portner K, Weilenmann C, Ma P *et al.* Opto-electronic memristors: Prospects and challenges in neuromorphic computing. *Appl Phys Lett* 2020; **117**: 230502.
86 Kim SY, Zhang H, Rivera-Sierra G, Fenollosa R, Rubio-Magnieto J, Bisquert J. Introduction to neuromorphic functions of memristors: The inductive nature of synapse potentiation. *J Appl Phys* 2025; **137**: 111101.
87 Shastri BJ, Tait AN, Ferreira de Lima T, Pernice WHP, Bhaskaran H, Wright CD *et al.* Photonics for artificial intelligence and neuromorphic computing. *Nat Photonics 2021 152* 2021; **15**: 102–114.
88 Sun chen, Wade mark T, Lee Y, Orcutt JS, Alloatti L, Georgas michael S *et al.* Single-chip microprocessor that communicates directly using light. *nature.com* 2015. doi:10.1038/nature16454.

89 Reed GT. Silicon Photonics: The State of the Art. *Silicon Photonics State Art* 2008; : 1–330.
90 Lanza M, Wong HSP, Pop E, Ielmini D, Strukov D, Regan BC *et al.* Recommended Methods to Study Resistive Switching Devices. *Adv Electron Mater* 2019; **5**: 1800143.
91 Alexoudi T, Kanellos GT, Pleros N. Optical RAM and integrated optical memories: a survey. *Light Sci Appl 2020 91* 2020; **9**: 1–16.
92 Almeida VR, Xu Q, Barrios CA, Lipson M. Guiding and confining light in void nanostructure. *Opt Lett Vol 29, Issue 11, pp 1209-1211* 2004; **29**: 1209–1211.
93 Tan H, Liu G, Zhu X, Yang H, Chen B, Chen X *et al.* An optoelectronic resistive switching memory with integrated demodulating and arithmetic functions. *Adv Mater* 2015; **27**: 2797–2803.
94 Liu J, Liu Y, Bian Y, Zheng Z, Zhou T, Zhu J. Hybrid wedge plasmon polariton waveguide with good fabrication-error-tolerance for ultra-deep-subwavelength mode confinement. *Opt Express, Vol 19, Issue 23, pp 22417-22422* 2011; **19**: 22417–22422.
95 Jung Y, Han H, Sharma A, Jeong J, Parkin SSP, Poon JKS. Integrated Hybrid VO2–Silicon Optical Memory. *ACS Photonics* 2022; **9**: 217–223.
96 Li C, Yu H, Shu T, Zhang Y, Wen C, Cao H *et al.* PZT optical memristors. *Nat Commun 2025 161* 2025; **16**: 6340-.
97 Amin R, George JK, Sun S, Ferreira De Lima T, Tait AN, Khurgin JB *et al.* ITO-based electro-absorption modulator for photonic neural activation function. *APL Mater* 2019; **7**: 81112.
98 Kumar S, Kumar A, Mishra RD, Babu P, Pandey SK, Devi S *et al.* Multilevel Nanophotonic Resistive Switching in Ag-ITO-SiO2on Silicon with Enhanced Optical Storage Density. *J Light Technol* 2024. doi:10.1109/JLT.2024.3474775.
99 Farmakidis N, Youngblood N, Li X, Tan J, Swett JL, Cheng Z *et al.* Plasmonic nanogap enhanced phase-change devices with dual electrical-optical functionality. *Sci Adv* 2019; **5**. doi:10.1126/SCIADV.AAW2687;CTYPE:STRING:JOURNAL.
100 Pianelli A, Dhama R, Judek J, Mazur R, Caglayan H. Si-CMOS compatible epsilon-near-zero metamaterial for two-color ultrafast all-optical switching. *Commun Phys 2024 71* 2024; **7**: 164-.
101 Oulton RF, Sorger VJ, Zentgraf T, Ma RM, Gladden C, Dai L *et al.* Plasmon lasers at deep subwavelength scale. *Nature* 2009; **461**: 629–632.
102 Dionne JA, Sweatlock LA, Sheldon MT, Alivisatos AP, Atwater HA. Silicon-based plasmonics for on-chip photonics. *IEEE J Sel Top Quantum Electron* 2010; **16**: 295–306.
103 Zhang Y, Chou JB, Li J, Li H, Du Q, Yadav A *et al.* Broadband transparent optical phase change materials for high-performance nonvolatile photonics. *Nat Commun 2019 101* 2019; **10**: 4279-.
104 Li Y, Shen W, Mao J, Zheng J, Xia W, Cai B *et al.* Intrinsic Volatile Photonic Memristors for Optical Neuromorphic Computing. *Laser Photonics Rev* 2026; : e03238.
105 Kazanskiy NL, Butt MA, Khonina SN. Optical Computing: Status and Perspectives. *Nanomater 2022, Vol 12, Page 2171* 2022; **12**: 2171.
106 Wang R, Zhang W, Wang S, Zeng T, Ma X, Wang H *et al.* Memristor-Based Signal Processing for Compressed Sensing. *Nanomater 2023, Vol 13, Page 1354* 2023; **13**: 1354.
107 Tuma T, Pantazi A, Le Gallo M, Sebastian A, Eleftheriou E. Stochastic phase-change neurons. *Nat Nanotechnol 2016 118* 2016; **11**: 693–699.
108 Tossoun B, Xiao X, Cheung S, Yuan Y, Peng Y, Srinivasan S *et al.* Large-Scale Integrated Photonic Device Platform for Energy-Efficient AI/ML Accelerators. *IEEE J Sel Top Quantum Electron* 2025; **31**. doi:10.1109/JSTQE.2025.3527904.
109 Atwater HA, Polman A. Plasmonics for improved photovoltaic devices. *Nat Mater* 2010; **9**: 205–213.
110 Hoessbacher C, Emboras A, Fedoryshyn Y, Kohl M, Melikyan A, Hillerkuss D *et al.* The plasmonic memristor: a latching optical switch. *Opt Vol 1, Issue 4, pp 198-202* 2014; **1**: 198–202.
111 Pandey SK, Mishra RD, Babu P, Mohanta N, Kumar S, Kumar M. Plasmonic Absorber Based on Engineered Cu-ITO Structure on Silicon With Low Voltage Tuning and High Extinction Ratio. *J Light Technol* 2024; **42**: 3779–3785.
112 Tossoun B, Sheng X, Strachan JP, Liang D. The memristor laser. *Tech Dig - Int Electron Devices Meet IEDM* 2020; **2020-December**: 7.6.1-7.6.4.
113 Cheung S, Tossoun B, Liang DI, Yuan Y, Hu Y, Kurczveil G *et al.* A Non-Volatile Heterogeneous Quantum Dot III-V/Si DFB Laser with Optical Memristive Behavior. 2026.https://arxiv.org/pdf/2605.22577 (accessed 22 May2026).
114 Papaioannou S, Kalavrouziotis D, Vyrsokinos K, Weeber JC, Hassan K, Markey L *et al.* Active plasmonics in WDM traffic switching applications. *Sci Rep* 2012; **2**. doi:10.1038/SREP00652.
115 Sorger VJ, Lanzillotti-Kimura ND, Ma RM, Zhang X. Ultra-compact silicon nanophotonic modulator with broadband response. *Nanophotonics* 2012; **1**: 17–22.
116 Nguyen HC, Hashimoto S, Shinkawa M, Baba T, Thomson DJ, Gardes FY *et al.* Compact and fast photonic crystal  silicon optical modulators. *Opt Express, Vol 20, Issue 20, pp 22465-22474* 2012; **20**: 22465–22474.
117 Joushaghani A, Kruger BA, Paradis S, Alain D, Stewart Aitchison J, Poon JKS. Sub-volt broadband hybrid plasmonic-vanadium dioxide switches. *Appl Phys Lett* 2013; **102**. doi:10.1063/1.4790834.
118 Lee HW, Papadakis G, Burgos SP, Chander K, Kriesch A, Pala R *et al.* Nanoscale Conducting Oxide PlasMOStor. *Nano Lett* 2014; **14**: 6463–6468.
119 Watts MR, Zortman WA, Trotter DC, Young RW, Lentine AL, Atkins DE *et al.* Vertical junction silicon microdisk modulators and switches. *Opt Express, Vol 19, Issue 22, pp 21989-22003* 2011; **19**: 21989–22003.
120 Liu J, Beals M, Pomerene A, Bernardis S, Sun R, Cheng J *et al.* Waveguide-integrated, ultralow-energy GeSi electro-absorption modulators. *Nat Photonics 2008 27* 2008; **2**: 433–437.
121 Geler-Kremer J, Eltes F, Stark P, Stark D, Caimi D, Siegwart H *et al.* A ferroelectric multilevel non-volatile photonic phase shifter. *Nat Photonics 2022 167* 2022; **16**: 491–497.
122 Stegmaier M, Ríos C, Bhaskaran H, David Wright C, P Pernice WH, Stegmaier M *et al.* Nonvolatile All-Optical 1 × 2 Switch for Chipscale Photonic Networks. *Adv Opt Mater* 2017; **5**: 1600346.
123 Fang Z, Chen R, Zheng J, Khan AI, Neilson KM, Geiger SJ *et al.* Ultra-low-energy programmable non-volatile silicon

photonics based on phase-change materials with graphene heaters. *Nat Nanotechnol 2022 178* 2022; **17**: 842–848.
124 Olivares I, Parra J, Sanchis P. Non-Volatile Photonic Memory Based on a SAHAS Configuration. *IEEE Photonics J* 2021; **13**. doi:10.1109/JPHOT.2021.3060144.
125 Song JF, Luo XS, Lim AEJ, Li C, Fang Q, Liow TY *et al.* Integrated photonics with programmable non-volatile memory. *Sci Reports 2016 61* 2016; **6**: 22616-.
126 Harris NC, Ma Y, Mower J, Baehr-Jones T, Englund D, Hochberg M *et al.* Efficient, compact and low loss thermo-optic phase shifter in silicon. *Opt Express, Vol 22, Issue 9, pp 10487-10493* 2014; **22**: 10487–10493.
127 Watts MR, Sun J, DeRose C, Trotter DC, Young RW, Nielson GN. Adiabatic thermo-optic Mach–Zehnder switch. *Opt Lett Vol 38, Issue 5, pp 733-735* 2013; **38**: 733–735.
128 Yu D, Liu LF, Chen B, Zhang FF, Gao B, Fu YH *et al.* Multilevel resistive switching characteristics in Ag/SiO 2/Pt RRAM devices. *2011 IEEE Int Conf Electron Devices Solid-State Circuits, EDSSC 2011* 2011. doi:10.1109/EDSSC.2011.6117721.
129 Wang LW, Huang CW, Lee KJ, Chu SY, Wang YH. Multi-Level Resistive Al/Ga2O3/ITO Switching Devices with Interlayers of Graphene Oxide for Neuromorphic Computing. *Nanomater 2023, Vol 13, Page 1851* 2023; **13**: 1851.
130 Haffner C, Heni W, Fedoryshyn Y, Elder DL, Melikyan A, Baeuerle B *et al.* High-speed plasmonic Mach-Zehnder modulator in a waveguide. *Eur Conf Opt Commun ECOC* 2014. doi:10.1109/ECOC.2014.6964271.
131 Battal E, Ozcan A, Okyay AK. Resistive switching-based electro-optical modulation. *Adv Opt Mater* 2014; **2**: 1149–1154.
132 Yuan Y, Peng Y, Cheung S, Sorin W V., Hooten S, Huang Z *et al.* All-silicon non-volatile optical memory based on photon avalanche-induced trapping. *Commun Phys 2025 81* 2025; **8**: 39-.
133 He C, Li J, Wu X, Chen P, Zhao J, Yin K *et al.* Tunable electroluminescence in planar graphene/sio2 memristors. *Adv Mater* 2013; **25**: 5593–5598.
134 Fang Y, Li Q, Wang T, Meng J, Sun QQ, Zhang DW *et al.* Photon-Memristive Device for Neuromorphic Computing. *Proc 2022 IEEE Int Conf Integr Circuits, Technol Appl ICTA 2022* 2022; : 50–51.
135 Zhu Z, Guglielmo G Di, Cheng Q, Glick M, Kwon J, Guan H *et al.* Photonic Switched Optically Connected Memory: An Approach to Address Memory Challenges in Deep Learning. *J Light Technol* 2020; **38**: 2814–2824.
136 Fraunhofer HHI technology in an all-optical RAM cell - PIC Magazine News. https://picmagazine.net/article/108218/Fraunhofer_HHI_technology_in_an_all-optical_RAM_cellnews (accessed 22 May2026).
137 Dai D, Fu X, Shi Y, He S. Experimental demonstration of an ultracompact Si-nanowire-based reflective arrayed-waveguide grating (de)multiplexer with photonic crystal reflectors. *Opt Lett* 2010; **35**: 2594–2596.
138 Cheung S, Su T, Okamoto K, Yoo SJB. Ultra-Compact Silicon Photonic 512 × 512 25 GHz Arrayed Waveguide Grating Router. *IEEE J Sel Top Quantum Electron* 2014; **20**. doi:10.1109/JSTQE.2013.2295879.
139 Youngblood N, Ríos Ocampo CA, Pernice WHP, Bhaskaran H. Integrated optical memristors. *Nat Photonics* 2023. doi:10.1038/S41566-023-01217-W.
140 Pi S, Li C, Jiang H, Xia W, Xin H, Yang JJ *et al.* Memristor crossbar arrays with 6-nm half-pitch and 2-nm critical dimension. *Nat Nanotechnol 2018 141* 2018; **14**: 35–39.
141 Wu C, Yu H, Lee S, Peng R, Takeuchi I, Li M. Programmable phase-change metasurfaces on waveguides for multimode photonic convolutional neural network. *Nat Commun 2021 121* 2021; **12**: 96-.
142 Liu W, Huang Y, Sun R, Fu T, Yang S, Chen H. Ultra-compact multi-task processor based on in-memory optical computing. *Light Sci Appl 2025 141* 2025; **14**: 134-.
143 Bandyopadhyay S, Sludds A, Krastanov S, Hamerly R, Harris N, Bunandar D *et al.* Single-chip photonic deep neural network with forward-only training. *Nat Photonics 2024 1812* 2024; **18**: 1335–1343.
144 Cheung S, London Y, Yuan Y, Tossoun B, Peng Y, Hu Y *et al.* Heterogeneous III-V/Si micro-ring laser array with multi-state non-volatile memory for ternary content-addressable memories. *Nat Commun 2025 161* 2025; **16**: 5020-.
145 Moschos T, Pappas C, Kovaios S, Roumpos I, Prapas A, Tsakyridis A *et al.* Nonlinear optical vector processing using linear silicon photonic circuits for 50 Gb/s memory and string similarity functions. *Nat Commun 2025 161* 2025; **16**: 11416-.
146 London Y, Van Vaerenbergh T, Ramini L, Descos A, Buonanno L, Youn J *et al.* Multiplexing in photonics as a resource for optical ternary content-addressable memory functionality. *Nanophotonics* 2023; **12**: 4147–4165.
147 Abel S, Eltes F, Ortmann JE, Messner A, Castera P, Wagner T *et al.* Large Pockels effect in micro- and nanostructured barium titanate integrated on silicon. *Nat Mater 2018 181* 2018; **18**: 42–47.
148 Rabiei P, Ma J, Khan S, Chiles J, Fathpour S, Fang AW *et al.* Heterogeneous lithium niobate photonics on silicon substrates. *Opt Express, Vol 21, Issue 21, pp 25573-25581* 2013; **21**: 25573–25581.
149 He M, Xu M, Ren Y, Jian J, Ruan Z, Xu Y *et al.* High-performance hybrid silicon and lithium niobate Mach–Zehnder modulators for 100 Gbit s−1 and beyond. *Nat Photonics 2019 135* 2019; **13**: 359–364.
150 Pintus P, Dumont M, Shah V, Murai T, Shoji Y, Huang D *et al.* Integrated non-reciprocal magneto-optics with ultra-high endurance for photonic in-memory computing. *Nat Photonics 2024 191* 2024; **19**: 54–62.
151 Bi L, Hu J, Jiang P, Kim DH, Dionne GF, Kimerling LC *et al.* On-chip optical isolation in monolithically integrated non-reciprocal optical resonators. *Nat Photonics 2011 512* 2011; **5**: 758–762.
152 Reed GT, Mashanovich G, Gardes FY, Thomson DJ. Silicon optical modulators. *Nat Photonics* 2010; **4**: 518–526.
153 Miller DAB. Attojoule Optoelectronics for Low-Energy Information Processing and Communications: a Tutorial Review. *J Light Technol* 2017; **35**: 346–396.
154 Yang F, Sun H, Zhang X, Chen D, Chen J, Zhang X *et al.* Optoelectronic Synaptic Memristor with Coupled Reversible Self-Rectifying and Bipolar Resistive Switching for Multifunctional Neuromorphic Applications. *Adv Funct Mater* 2025; **36**: e16894.
155 Kantor B, Kantor B, Kantor B, Ackermann L, Ackermann L, Ackermann L *et al.* Individual nanostructures in an epsilon-near-zero material probed with 3D-sculpted light. *Opt Express, Vol 32, Issue 27, pp 47800-47809* 2024; **32**: 47800–

47809.
156 Liang D, Srinivasan S, Kurczveil G, Tossoun B, Cheung S, Yuan Y *et al.* An Energy-Efficient and Bandwidth-Scalable DWDM Heterogeneous Silicon Photonics Integration Platform. *IEEE J Sel Top Quantum Electron* 2022; **28**. doi:10.1109/JSTQE.2022.3181939.
157 Yesilkoy F, Arvelo ER, Jahani Y, Liu M, Tittl A, Cevher V *et al.* Ultrasensitive hyperspectral imaging and biodetection enabled by dielectric metasurfaces. *Nat Photonics 2019 136* 2019; **13**: 390–396.
158 De Vos K, Bartolozzi I, Schacht E, Bienstman P, Baets R, Yalçin A *et al.* Silicon-on-Insulator microring resonator for sensitive and label-free biosensing. *Opt Express, Vol 15, Issue 12, pp 7610-7615* 2007; **15**: 7610–7615.
159 Fan X, White IM, Shopova SI, Zhu H, Suter JD, Sun Y. Sensitive optical biosensors for unlabeled targets: A review. *Anal Chim Acta* 2008; **620**: 8–26.
160 Ielmini D, Wong HSP. In-memory computing with resistive switching devices. *Nat Electron 2018 16* 2018; **1**: 333–343.
161 Zahoor F, Hussin FA, Isyaku UB, Gupta S, Khanday FA, Chattopadhyay A *et al.* Resistive random access memory: introduction to device mechanism, materials and application to neuromorphic computing. *Discov Nano 2023 181* 2023; **18**: 36-.
162 Zhang R, He Y, Zhang Y, An S, Zhu Q, Li X *et al.* Ultracompact and low-power-consumption silicon thermo-optic switch for high-speed data. *Nanophotonics* 2020; **10**: 937–945.
163 Alfaraj N, Helmy AS. Silicon-Integrated Next-Generation Plasmonic Devices for Energy-Efficient Semiconductor Applications. *Adv Mater Technol* 2025; **10**: e00389.
164 Tossoun B, Liang D, Cheung S, Fang Z, Sheng X, Strachan JP *et al.* High-speed and energy-efficient non-volatile silicon photonic memory based on heterogeneously integrated memresonator. *Nat Commun 2024 151* 2024; **15**: 551-.
165 Zhang EJ, Orcutt JS, Tombez L, Kamlapurkar S, Green WMJ. Methane absorption spectroscopy on a silicon photonic chip. *Opt Vol 4, Issue 11, pp 1322-1325* 2017; **4**: 1322–1325.